\documentclass[12pt]{article}
\usepackage{epsfig,amsfonts,amssymb,amsbsy}
\def\cvp{\raise 2pt\hbox{,}}

\def\tr{\mathop{\rm tr}\nolimits}
\def\det{\mathop{\rm det}\nolimits}
\def\im{\mathop{\rm Im}\nolimits}
\def\re{\mathop{\rm Re}\nolimits}

\def\partialsl{\partial\llap{/\hskip .2pt}}
\def\nablasl{\nabla\llap{/\hskip 2.4pt}}
\def\plb#1#2#3{{\it Phys.\ Lett.\ }{\bf B #1} (#2) #3}
\def\npb#1#2#3{{\it Nucl.\ Phys.\ }{\bf B #1} (#2) #3}
\def\prl#1#2#3{{\it Phys.\ Rev.\ Lett.\ }{\bf #1} (#2) #3}
\def\jhep#1#2#3{{\it J. High Energy Phys.\ }{\bf #1} (#2) #3}
\def\prd#1#2#3{{\it Phys.\ Rev.\ }{\bf D #1} (#2) #3}
\def\atmp#1#2#3{{\it Adv.\ Theor.\ Math.\ Phys.\ }{\bf #1} (#2) #3}
\def\cmp#1#2#3{{\it Comm.\ Math.\ Phys.\ }{\bf #1} (#2) #3}
\def\pr#1#2#3{{\it Phys.\ Rep.\ }{\bf #1} (#2) #3}

\def\suN{{\rm SU}(N)}
\def\cpN{{\mathbb C}P^{N}}

\def\tp{\theta_{+}}\def\tm{\theta_{-}}\def\tbp{\bar\theta_{+}}
\def\tbm{\bar\theta_{-}}
\def\weff{\tilde W_{\rm eff}}
\begin{document}
%
%
\pagestyle{empty}
{\parskip 0in
\hfill NEIP-01-008

\hfill PUPT-1997

\hfill LPTENS-01/11

\hfill hep-th/0202002}

\vfill
\begin{center}
{\LARGE Large $N$ and double scaling limits in two dimensions}

\vspace{0.4in}

Frank F{\scshape errari}{\renewcommand{\thefootnote}{$\!\!\dagger$}
\footnote{On leave of absence from Centre 
National de la Recherche Scientifique, Laboratoire de Physique 
Th\'eorique de l'\'Ecole Normale Sup\'erieure, Paris, France.}}\\
\medskip
{\it Institut de Physique, Universit\'e de Neuch\^atel\\
rue A.-L.~Br\'eguet 1, CH-2000 Neuch\^atel, Switzerland\\
and\\
Joseph Henry Laboratories\\
Princeton University, Princeton, New Jersey 08544, USA}\\
\smallskip
{\tt frank.ferrari@unine.ch}
\end{center}
\vfill\noindent
Recently, the author has constructed a series of four dimensional 
non-critical string theories with eight supercharges, dual to 
theories of light electric and magnetic charges,
for which exact formulas for the central charge of the space-time 
supersymmetry algebra as a function of the world-sheet couplings were 
obtained. The basic idea was to generalize the old matrix model 
approach, replacing the simple matrix integrals by the four 
dimensional matrix path integrals of ${\cal N}=2$ supersymmetric 
Yang-Mills theory, and the Kazakov critical points by the 
Argyres-Douglas critical points. In the present paper, we study 
qualitatively similar toy path integrals corresponding to the two 
dimensional ${\cal N}=2$ supersymmetric non-linear $\sigma$ model with 
target space $\cpN$ and twisted mass terms. This theory has some very 
strong similarities with ${\cal N}=2$ super Yang-Mills, including the 
presence of critical points in the vicinity of which the large $N$ 
expansion is IR divergent. The model being exactly solvable at large 
$N$, we can study non-BPS observables and give full proofs that double 
scaling limits exist and correspond to universal continuum limits. A 
complete characterization of the double scaled theories is given. We 
find evidence for dimensional transmutation of the string coupling in 
some non-critical string theories. 
We also identify en passant some non-BPS particles that
become massless at the singularities in addition to the usual BPS states.
\vfill
\begin{flushleft}
January 2002
\end{flushleft}
\newpage\pagestyle{plain}
\baselineskip 16pt
\setcounter{footnote}{0}
\renewcommand{\theequation}{\thesection.\arabic{equation}}
%
\section{Introduction}
In two recent papers \cite{fer1,fer2}, unexpected
properties of the large $N$ limit of ${\cal N}=2$ super 
Yang-Mills theory with gauge group $\suN$
have been discovered and exploited. The r\^ole of instantons
at strong coupling, which has always been elusive, 
has been elucidated in this context by computing the 
large $N$ expansion of BPS observables. It 
turns out \cite{fer1} that the large instantons disintegrate into 
`fractional instantons' which give non-trivial contributions at each 
order in $1/N$. These `fractional instantons' 
are thus in particular responsible for the 
presence of open strings in the string theory dual \cite{fer1}, in 
addition to the familiar closed strings contributing at each order in 
$1/N^{2}$ \cite{tHooft}. The 
fractional instanton series have a finite radius of convergence, and 
they diverge precisely at the singularities on moduli 
space. This breakdown of the large $N$ expansion is interpreted 
\cite{fer1} as coming from infrared divergences due to the presence 
of a critical point. Similar divergences were 
encountered long ago in the study of
the large $N$ limit of ordinary zero dimensional
matrix integrals \cite{IZ} near critical points \cite{DK}. In those 
simple cases, it was shown in \cite{BK} that the divergences can be 
used to define universal double scaling limits from which one can 
extract exact results in continuum string theories. The critical string 
theories were defined in less than two space-time dimensions (the 
$c=1$ barrier) because the only tractable cases were zero 
or one dimensional path 
integrals. This limitation was overcome in \cite{fer2},
where it was argued that the divergences found in \cite{fer1} can also 
be used to define double scaling limits, yielding exact results in
four dimensional non-critical (or five dimensional critical) string 
theories. The string theories so obtained are dual to theories of 
light electric and magnetic charges which do not have any obvious 
description in terms of a local lagrangian quantum field theory.

The four dimensional results of
\cite{fer1,fer2} were derived by studying the large $N$ expansion of the 
Seiberg-Witten period integrals \cite{SW,sun}.
These periods give the central 
charge $Z$ of the supersymmetry algebra as a function of the moduli, 
and thus the masses of BPS states
\begin{equation}
\label{MBPS}
M_{\rm BPS} = \sqrt{2}\, |Z|.
\end{equation}
This class of observables is ideal to deduce the strong coupling
behaviour of instantons because the perturbative quantum corrections 
stop at one loop and all the non-trivial contributions can be 
understood as coming exclusively from instantons. For our purposes, 
however, the consideration of those special protected amplitudes is 
not enough. To give a full proof of the existence of
double scaling limits, one must study in principle {\it all}
the observables, including those with a non-trivial perturbative 
expansion, or equivalently the full path integral.
Moreover, the heuristic picture for the appearance of
a continuum string theory in the limit relies on the observation
that very large Feynman graphs dominate near the critical points 
\cite{DK}, a fact that can in principle
be checked on generic amplitudes but obviously
not on the BPS observables for which perturbation theory is 
trivial. A related point is to understand the 
universality of the double scaling limits. One argument for universality, 
put forward in \cite{fer2}, is that the double scaling limits are always 
low energy limits of the original field theory. It was observed, however, 
that to a given CFT in the infrared can be associated two different double 
scaled theories (first class or second class singularities in the 
terminology of \cite{fer1}). It would thus clearly be desirable to have a 
more precise characterization of the string theories obtained in the
scaling limits.

The purpose of the present paper is to shed some light on all the 
above issues by studying a particular two dimensional model 
which is a very close relative to ${\cal N}=2$ super Yang-Mills in 
four dimensions. The model has an exactly calculable central charge 
with the same non-renormalization theorems as in four dimensions and 
the same BPS mass formula (\ref{MBPS}). The analysis of \cite{fer1,fer2} 
can thus be reproduced, with, as we will demonstrate, qualitatively the same
results (appearance of fractional instantons, breakdown of the 
large $N$ expansion at critical points, possibility to define double 
scaling limits for which exactly known BPS amplitudes
have a finite limit). Moreover, and this is our main point,
as our two-dimensional model is
solvable in the large $N$ limit, we can go far beyond the analysis of 
the BPS observables. We will actually be able to give full 
proofs of the existence of the double scaling limits, and we will 
give a complete characterization of the double scaled theories.

The two dimensional theory we consider
is the supersymmetric non-linear $\sigma$ model with target space 
$\cpN$ and ${\cal N}=2$ preserving mass terms for the would be 
Goldstone bosons. The close relationship of this model with super 
Yang-Mills was emphasized in \cite{Dorey}, and the general 
analogy between mass terms in non-linear $\sigma$ models and Higgs 
vevs in gauge theories was discussed at lenght in \cite{fer3}. It 
turns out that the two dimensional supersymmetric $\cpN$ model and the 
four dimensional super Yang-Mills theory are both asymptotically free 
with a dynamically generated mass scale, have instantons,
share the same types of 
non-renormalization theorems, and have both BPS solitonic states 
that can become massless at strong coupling singularities. The moduli 
space of the gauge theory is analogous to the space of mass parameters 
of the non-linear $\sigma$ model \cite{fer3}. This very strong analogy 
can even be made quantitative if one adds 
$N_{\rm f}=N$ matter hypermultiplets in the fundamental to the pure
$\suN$ theory, and choose the hypers masses $m_{i}$ to match the 
Higgs vevs $\phi_{i}$. One can then show 
\cite{Dorey} that the central charges of the four dimensional 
Yang-Mills theory and of the two dimensional non-linear $\sigma$ 
model are actually equal as functions of the $m_{i}$s,
\begin{equation}
\label{corr}
Z_{\rm YM}(\phi_{i},m_{i}=\phi_{i}) = Z_{\sigma\, {\rm model}}(m_{i}).
\end{equation}
The simplest way to understand this relation is to look at the 
respective brane constructions of the models
\cite{witbra,hr}. The central charges $Z$ are determined 
by the shape of Neveu-Schwarz five-branes which are
bent by D4 branes ending on them. It turns out that the relevant
configurations of branes 
are the same for the two theories, from which (\ref{corr}) follows. 
In \cite{Dorey2} it was argued that equation (\ref{corr}) also 
implies that the stable BPS states are the same in two and four 
dimensions. It is indeed known in the simplest case $N=2$ 
\cite{FB} that the knowledge of $Z$ goes a long way toward determining
the BPS spectra.

We have organized the paper as follows. In Section 2, we give a 
rather detailed presentation of various classic results about our two 
dimensional model, including the derivation of the central charge as 
a function of the masses $m_{i}$.
Our goal was to make the paper as self-contained as possible.
Taking for granted the formulas (\ref{Zeq}) and (\ref{wex}),
the reader may wish to proceed directly to Section 3 where the 
large $N$ limit of the BPS mass formula is analysed, and the double 
scaling limits are defined. Section 3 does in two dimensions precisely
what was done in \cite{fer1,fer2} in four dimensions, and we recover 
the same qualitative physics (enhan\c con, fractional instantons, IR 
divergences, double scaled amplitudes). We also discuss at an 
elementary level the universality of the double scaled theories.
In Section 4, we give a general 
analysis of the large $N$ expansion. We discuss in details the 
physical significance of the double scaling limits, first in a 
heuristic way by using the dual Feynman graphs representation, then 
rigorously by using the exact solution of our model at large $N$.
The main outcome is a 
full proof of the existence and universality of the double scaling
limits exhibited in Section 3. We show that the `string' coupling 
undergoes dimensional transmutation for first class theories.
Another interesting finding is that
BPS/anti-BPS bound states can become massless at singularities, in 
addition to the standard BPS solitons. We then briefly comment, 
in Section 5,
on other models with ${\cal N}=1$ or ${\cal N}=0$ supersymmetry, and 
we conclude in Section 6 by giving possible future directions of research.

\section{Classic results}
\setcounter{equation}{0}
\subsection{Lagrangian, symmetries and renormalization}
\subsubsection{${\cal N}=2$ superspace and superfields}
The ${\cal N}=2$ superspace in two dimensions is the dimensional 
reduction of the standard ${\cal N}=1$ superspace in four dimensions, 
with anticommuting coordinates $\theta_{\pm}$ and $\bar\theta_{\pm}$, 
supersymmetry generators
\begin{equation}
\label{Qdef}
Q_{\pm}=\mp i {\partial\over\partial\theta_{\mp}}\pm 
2\bar\theta_{\mp}\partial_{\pm}\, ,\quad 
\bar Q_{\pm} = \pm i {\partial\over\partial\bar\theta_{\mp}} \mp 
2\theta_{\mp}\partial_{\pm}\, ,
\end{equation}
and supercovariant derivatives
\begin{equation}
\label{Ddef}
D_{\pm}=\mp {\partial\over\partial\theta_{\mp}} \pm 2i 
\bar\theta_{\mp} \partial_{\pm}\, ,\quad
\bar D_{\pm} = \pm {\partial\over\partial\bar\theta_{\mp}} \mp 
2i\theta_{\mp}\partial_{\pm}\, .
\end{equation}
The two dimensional case has some important peculiarities due to the 
fact that Lorentz transformations do not mix the right and left moving 
components. One can define two R charges, the ordinary fermion number 
${\rm U}(1)_{\rm F}$ under which $Q_{-}$ and $Q_{+}$ have a charge 
$+1$, and an axial ${\rm U}(1)_{\rm A}$ under which $Q_{-}$ and $Q_{+}$ 
have respectively a charge $+1$ and $-1$.
Similarly, in addition to ordinary chiral 
superfields $\Phi$ defined by the equations
\begin{equation}
\label{chiraldef}
\bar D_{+}\Phi = \bar D_{-}\Phi =0\, ,
\end{equation}
one can define twisted chiral superfields $\Sigma$ by the equations
\begin{equation}
\label{twistdef}
\bar D_{+}\Sigma = D_{-}\Sigma = 0\, .
\end{equation}
Ordinary and twisted chiral superfields are exchanged by mirror 
symmetry.
Gauge fields corresponding to gauge symmetries acting on chiral 
superfields belong to vector superfields $V=V^{\dagger}$ whose 
field strengths turn out to be twisted chiral 
superfields $\Sigma$ defined by
\begin{equation}
\label{sigmadef}
\Sigma = \bar D_{+}D_{-}V\, .
\end{equation}
The relation (\ref{twistdef}) and gauge invariance are demonstrated 
by using the two dimensional formula
\begin{equation}
\label{alg1}
\{ \bar D_{+},D_{-} \} =0\, .
\end{equation}
In components, the various superfields can be decomposed by using the 
variables $y^{\pm} = x^{\pm} -2i \theta_{\mp}\bar\theta_{\mp}$,
$\tilde y^{\pm} = x^{\pm} \mp 2 i \theta_{\mp}\bar\theta_{\mp}$, and 
the Wess-Zumino gauge for $V$,
\begin{eqnarray}
\label{superfields}
\Phi (x,\theta,\bar\theta) &=& \phi (y) + 
\sqrt{2}\theta_{+}\psi_{-} - \sqrt{2} \theta_{-}\psi_{+}+ 
2\theta_{+}\theta_{-}F\, ,\\
V(x,\theta,\bar\theta) &=& -\theta_{+}\bar\theta_{+}v^{+}(x) - 
\theta_{-}\bar\theta_{-}v^{-} + \theta_{+}\bar\theta_{-}\sigma  
+\theta_{-}\bar\theta_{+}\sigma^{\dagger}
+ 2 i \theta_{-}\theta_{+} 
(\bar\theta_{+}\bar\lambda_{-}-\bar\theta_{-}\bar\lambda_{+})\nonumber\\
&&\hskip 3.5cm + 2i \bar\theta_{+}\bar\theta_{-} ( \theta_{+}\lambda_{-} - 
\theta_{-}\lambda_{+}) + 2 
\theta_{-}\theta_{+}\bar\theta_{+}\bar\theta_{-} D\, ,\\
\Sigma(x,\theta,\bar\theta) &=& \sigma (\tilde y) -2i 
(\tm\bar\lambda_{+}+\tbp\lambda_{-}) + 2\tbp\tm (D - i v)\, .\\
\nonumber
\end{eqnarray}
We note that the field strength 
$v=v_{01}=\partial_{0}v_{1}-\partial_{1}v_{0}$ is an auxiliary field, 
a result consistent with the fact that gauge fields do not propagate 
in two dimensions.
\subsubsection{Lagrangian}
The most general manifestly supersymmetric and renormalizable 
lagrangian can be written as a sum of $D$-, $F$- and 
twisted $F$-terms,
\begin{equation}
\label{Lgen}
L = {1\over 4}\int\! d^{4}\theta\, K(\Phi,\Phi^{\dagger},
\Sigma,\Sigma^{\dagger},V)
- \re\int\! d\tm d\tp\, W_{\rm cl}(\Phi) - \re\int\! d\tm d\tbp\, \tilde 
W_{\rm cl}(\Sigma)\, ,
\end{equation}
where the K\"ahler potential $K$ is an arbitrary real function and
$W_{\rm cl}$ and $\tilde W_{\rm cl}$ are holomorphic 
functions called the superpotential and twisted superpotential 
respectively. The measure of integration over the whole of superspace 
is defined to be $d^{4}\theta = d\tp d\tm d\tbm d\tbp$.
The supersymmetric $\cpN$ model is 
defined in terms of $N$ chiral superfields $Z_{i}$ locally
parametrizing the complex K\"ahler manifold $\cpN$ with K\"ahler potential
\begin{equation}
\label{KP}
K = {4\pi\over g^{2}}\, \ln \Bigl( 1+\sum_{i=1}^{N}Z_{i}^{\dagger}Z_{i}
\Bigr)\, .
\end{equation}
When the coupling $g$ is small, the target space manifold is large and 
vice-versa. To describe $\cpN$ globally, we actually need $N+1$ 
coordinate patches $Z_{i}^{(j)}$, $1\leq i,j \leq N+1$, $i\not = j$, 
related to each other by $Z_{i}^{(j)}= Z_{i}^{(k)}/Z_{j}^{(k)}$.
A more elegant description of $\cpN$ is in terms of $N+1$
complex variables $\phi$ constrained by
\begin{equation}
\label{cpNeq}
\sum_{i=1}^{N+1} |\phi_{i}|^{2} = {4\pi\over g^{2}}
\end{equation}
and with the ${\rm U}(1)$ identification
\begin{equation}
\label{cpNid}
\phi_{i} \sim e^{i\alpha}\phi_{i}\, .
\end{equation}
The coordinates $Z_{i}^{(j)}=\phi_{i}/\phi_{j}$ can be used
as long as $\phi_{j}\not =0$.
By introducing chiral superfields $\Phi_{i} = \phi_{i} + \cdots$, 
a Lagrange multiplier vector superfield $V$ and associated $\Sigma = 
\bar D_{+} D_{-}V$, and a twisted superpotential
\begin{equation}
\label{Wcl}
\tilde W_{\rm cl} = -{i\over 2}\tau\Sigma =- {i\over 2} \left( 
{\theta\over 2\pi} + i {4\pi\over g^{2}}\right)\Sigma ,
\end{equation}
the lagrangian can be written as
\begin{equation}
\label{LcpN}
L = {1\over 4}\int\! d^{4}\theta \sum_{i=1}^{N+1} 
\Phi_{i}^{\dagger} e^{2V}\Phi_{i} - \re\int\! d\tm d\tbp \tilde 
W_{\rm cl}(\Sigma)\, .
\end{equation}
The vector superfield $V$
implement the gauge symmetry (\ref{cpNid}) and the twisted superpotential
implement the constraint (\ref{cpNeq}). The $\theta$ angle term 
corresponds to 
the total derivative $\theta v/(2\pi)$ in the lagrangian. Such a 
term is actually important at weak coupling because of the presence of 
instantons, for which
\begin{equation}
\label{instq}
\int\! {v\over 2\pi} \in {\mathbb Z}\, .
\end{equation}
The $\theta$ term plays a r\^ole at strong coupling as well, as we will
explain below. By integrating out $V$ from (\ref{LcpN}), we recover 
the pure $D$-term lagrangian with K\"ahler potential (\ref{KP}) plus
the topological $\theta$ angle term  proportional to
$\theta\epsilon^{\mu\nu} 
\partial_{i}\bar\partial_{j}K \partial_{\mu}z_{i}\partial_{\nu}\bar z_{j}$.

As discussed in the introduction, we want to introduce ${\cal N}=2$ 
preserving mass terms for the would-be Goldstone bosons $Z_{i}$. 
These masses play the same qualitative r\^ole as Higgs
vevs in gauge theories. There is no suitable manifestly 
supersymmetric mass
term, because a superpotential must be holomorphic and  
there is no non-trivial holomorphic function on a compact complex manifold. 
However, as first discussed in \cite{AGmass}, non-trivial ${\cal 
N}=2$ preserving mass terms associated with holomorphic isometries of 
the target space manifold can be written down. An important property 
of such terms is that they induce a 
non-trivial contribution to the central charge of the supersymmetry 
algebra. In our case, there are 
$N+1$ holomorphic Killing vectors associated with the $N+1$ symmetries
\begin{equation}
\label{u1i}
{\rm U}(1)_{i}:\ \phi_{j}\longmapsto 
e^{i\alpha_{i}\delta_{ij}}\phi_{j}\, ,
\end{equation}
$N$ of which are independent taking into account the identification
(\ref{cpNid}). We thus get $N$ independent masses for the $N$ 
fields $Z_{i}$. The explicit form of these terms can be obtained 
by gauging the ${\rm U}(1)_{i}$ symmetries, which amounts to 
replacing $\Phi_{i}^{\dagger} e^{2V}\Phi_{i}$ in (\ref{LcpN}) 
by $\Phi_{i}e^{2(V+V_{i})}\Phi_{i}$, and then by freezing 
$\Sigma_{i}=m_{i}$. This procedure of gauging also explains why
the $m_{i}$s can be interpreted as the position of branes. To write 
down the form of the final lagrangian, which we will use in Section 
4, it is convenient to introduce $\gamma$ matrices and Dirac spinors,
\begin{eqnarray}
\label{diracnot}
&&\hskip 1.43cm
\gamma^{0}=-\sigma^{1}\, ,\quad \gamma^{1}=i\sigma^{2}\, ,\quad
\gamma^{3} = \gamma^{0}\gamma^{1}=\sigma^{3}\, ,\\
&&\hskip -1.5cm\psi = \pmatrix{\psi_{-}\cr\psi_{+}\cr} ,\quad \lambda = 
\pmatrix{\lambda_{-}\cr\lambda_{+}\cr} ,\quad {}^{\rm c}\lambda = 
\pmatrix{\bar\lambda_{-}\cr -\bar\lambda_{+}}=\mu\, ,\quad
\bar\psi = \psi^{\dagger}\gamma^{0}\, ,\quad\bar\mu = 
\mu^{\dagger}\gamma^{0}\, ,\\ \nonumber
\end{eqnarray}
in terms of which
\begin{eqnarray}
\label{lag}
&&\hskip -1.5cm L=\sum_{i=1}^{N+1}\Bigl( 
-(\partial_{\mu}-iv_{\mu})\phi_{i}^{\dagger}\, (\partial^{\mu}+iv^{\mu})
\phi_{i} - |\sigma + m_{i}|^{2}\, |\phi_{i}|^{2}\nonumber\\ 
&&\hskip .7cm +\bar\psi_{i}\left(
i\gamma^{\mu}\partial_{\mu}-\gamma^{\mu}v_{\mu}+\re 
(\sigma + m_{i}) -i \gamma^{3}\im (\sigma + m_{i})\right)\psi_{i}\Bigr)
\nonumber\\
&&\hskip .7cm  +{\theta\over 2\pi}\, v
+ D\,\Bigl( \sum_{i=1}^{N+1}|\phi_{i}|^{2} - {4\pi\over g^{2}}\Bigr) + 
i\sqrt{2}\,\bar\mu\sum_{i=1}^{N+1} \psi_{i}\phi_{i}^{\dagger} - 
i\sqrt{2}\sum_{i=1}^{N+1}\phi_{i}\bar\psi_{i}\,\mu\, .\\ \nonumber
\end{eqnarray}
We get the supersymmetric partner of the bosonic constraint 
(\ref{cpNeq}), $\sum_{i=1}^{N+1}\phi_{i}^{\dagger}\psi_{i}=0$, as 
well as the classical potential
\begin{equation}
\label{Vcl}
V_{\rm cl} = \sum_{i=1}^{N+1} |\sigma + m_{i}|^{2}\, |\phi_{i}|^{2}\, .
\end{equation}
The potential yields $N+1$ inequivalent vacua $|i\rangle$,
\begin{equation}
\label{vaccl}
\langle\sigma\rangle _{i,\rm cl} = -m_{i}\, ,\quad \langle 
|\phi_{j}|^{2}\rangle _{i,\rm cl} = {4\pi\over g^{2}}\, 
\delta_{ij}\, ,
\end{equation}
consistently with the Witten index $\tr (-1)^{F} = \chi_{\rm Euler}
(\cpN) = N+1$.
\subsubsection{Symmetries and non-renormalization theorem}
The classical pure $\cpN$ model has a ${\rm SU}(N+1)\times {\rm 
U}(1)_{\rm F} \times {\rm U}(1)_{\rm A}$ bosonic global symmetry in 
addition to ${\cal N}=2$ supersymmetry. The ${\rm U}(1)_{\rm A}$ 
symmetry is preserved by the twisted superpotential (\ref{Wcl})
by assigning ${\rm U}(1)_{\rm A}$ charge 2 to $\Sigma$. 
However, ${\rm U}(1)_{\rm A}$ acts chirally on the $N+1$ charged fermions,
$\psi_{i,\pm}\mapsto e^{\mp i\alpha}\psi_{i,\pm}$, and will thus be 
anomalous. The anomalous transformation determines exactly the gauge 
theoretic perturbative effective twisted superpotential
\begin{equation}
\label{W1l}
\tilde W_{\rm pert}(\Sigma ,m=0) = {N+1\over 4\pi}\, \Sigma\ln 
{\Sigma\over e\Lambda}\, \cvp
\end{equation}
where $\Lambda$ is the complexified dynamically generated scale of the 
theory,
\begin{equation}
\label{lcplex}
\Lambda^{N+1} = |\Lambda|^{N+1}e^{i\theta},
\end{equation}
with a convenient normalization. 
Similarly, the anomaly can be used to deduce the gauge theoretic
perturbative twisted superpotential for arbitrary twisted masses
$m_{i}$, by assigning charge 2 to the masses,
\begin{equation}
\label{Wpert}
\tilde W_{\rm pert}(\Sigma ,m) = {1\over 
4\pi}\sum_{i=1}^{N+1}(\Sigma + m_{i})\ln {\Sigma + 
m_{i}\over e\Lambda}\, \cdotp
\end{equation}
These formulas could have been deduced equivalently
by a direct computation of 
the quantum corrections to (\ref{Wcl}). This shows in particular 
that the running of the coupling $g$ is given by 
the $\sigma$ model one-loop contribution,
\begin{equation}
\label{beta}
{1\over g^{2}(\mu)} = {N+1\over 8\pi^{2}}\, \ln 
{\mu\over |\Lambda|}\, \cdotp
\end{equation}
Note that the ${\cal N}=2$, $N_{\rm c}=N_{\rm f}=N+1$ gauge theory in four 
dimensions has precisely the same running coupling \cite{Dorey}. 

The supersymmetry algebra
\begin{equation}
\label{susyalg}
\{ Q_{\pm},\bar Q_{\pm}\} = -4 P_{\pm}\, ,\quad
\{\bar Q_{+},Q_{-}\} = 2\sqrt{2}\, Z\, ,\quad \{Q_{+},Q_{-}\} =0\, ,
\end{equation}
implies a BPS bound on the one-particle states masses $M\geq \sqrt{2}\,Z$,
and BPS states are defined to saturate this bound (\ref{MBPS}). 
The central charge $Z$ is a linear combination of the charges $S_{i}$ 
associated with the ${\rm U}(1)_{i}$ transformations (\ref{u1i}) and 
the topological charges $T_{i}$ for solitons interpolating between vacua 
$|i\rangle$ and $|j\rangle$ (for which $T_{k}=\delta_{ik}-\delta_{jk}$),
\begin{equation}
\label{Zeq}
Z = i\sqrt{2}\, \sum_{i=1}^{N+1}T_{i} 
\tilde W_{\rm eff}(\langle\sigma\rangle 
_{i},m) + {1\over\sqrt{2}}\sum_{i=1}^{N+1} m_{i}S_{i}\, .
\end{equation}
$\tilde W_{\rm eff}$ is the 
exact effective twisted superpotential, whose classical and 
perturbative formulas are given respectively by (\ref{Wcl}) and 
(\ref{Wpert}). The $\langle\sigma\rangle _{i}$s satisfy the vacuum 
equation
\begin{equation}
\label{vides}
{\partial\tilde W_{\rm eff}\over \partial\sigma} 
(\sigma = \langle\sigma\rangle _{i},m) =0\, ,
\end{equation}
and are classically given by (\ref{vaccl}).
\subsection{The exact non-perturbative superpotential}
The superpotential (\ref{Wpert}) has been deduced from an anomaly 
calculation in the {\it gauge} theory (\ref{lag}). This gauge theory has an 
infinite gauge coupling $E$ since there is no kinetic term for the 
gauge field $\Sigma$, whereas the anomaly is computed in perturbation 
theory in $E$ (perturbation theory in $E$ is not to be confused with 
perturbation theory in the non-linear $\sigma$ model coupling $g$). 
The formula is nevertheless exact, up to an important 
subtlety discussed at the end of this Section. There are many ways 
to prove this result. One can use the brane construction to 
compute $\tilde W_{\rm eff}$ \cite{hr}. One can also use an improved 
Witten index \cite{CFIV} to show that $M_{\rm BPS}$, and thus the 
central charge $Z$ and $\tilde 
W_{\rm eff}$, do not depend on the $D$-terms and thus do not 
depend on $E$. The result (\ref{Wpert}), known to be valid when 
$E\rightarrow 0$, is thus also valid when $E\rightarrow\infty$.
The fact that the 
central charge $Z$ does not depend on $E$ can also be understood from 
Gauss's law, which imply that $Z$ can be computed from the behaviour 
of the fields at large distances, whereas $E$ is an irrelevant 
coupling in two dimensions. Yet another derivation, which is
both elementary and rigorous, is to note that the exact effective action 
for $\Sigma$ can be deduced by integrating out the fields $\Phi_{i}$ 
from (\ref{lag}). Since the $\Phi_{i}$s appear only quadratically, 
this can be done exactly. To isolate the 
twisted superpotential term from the general non-local effective 
action for $\Sigma$, one uses the fact that the four-momentum 
$P_{\mu}$ can be written as an anticommutator of the 
supercharges, and thus that any non-local F-term can also be written
as a $D$-term. The most general $F$-term is then given by
the local twisted superpotential. To compute $\tilde W_{\rm eff}$, it 
is thus enough to consider constant fields, and to set the fermions 
and the field strenght $v$ to zero. The general effective action then 
admits an expansion is powers of $D$,
\begin{equation}
\label{Dexp}
S_{\rm eff} = \int\! d^{2}x \Bigl( -2D \re {\partial\tilde W_{\rm 
eff}\over\partial\sigma } + {\cal O}(D^{2})\Bigr)\, .
\end{equation}
The $D$-terms can contribute only at order $D^{2}$ or higher. The term
linear in $D$, which is given by a simple one-loop calculation in the 
gauge theory, together with the analyticity properties of the twisted 
superpotential, yield the formula (\ref{Wpert}).

There is a difficulty with (\ref{Wpert}), which seems to 
be at the origin of some confusion in the literature. The formula 
is ambiguous because the logarithm is a multivalued function. The 
ambiguity corresponds to adding a term $ip\Sigma /2$, $p\in {\mathbb 
Z}$, to the twisted superpotential, or equivalently to shifting the 
$\theta$ angle by $-2\pi p$.
If we are in the vacuum $|i\rangle$, and if the masses 
are such that $|m_{j}-m_{i}|>>\Lambda$ for $i\not =j$, then the 
physics is weakly coupled and the structure of the vacuum is 
semiclassical. In particular, this means that the boundary conditions 
at infinity are such that the quantization condition (\ref{instq}) holds,
and thus the ambiguity on $\weff$ is unphysical. This
can actually be checked explicitly. The 
physical content of $\weff$ is summarized by the vacuum equation
\begin{equation}
\label{vaceq}
{\partial\weff\over\partial\sigma} = 0\, .
\end{equation}
Using (\ref{Wpert}), this reduces to
\begin{eqnarray}
&&\hskip -1cm\prod_{j=1}^{N+1}|\sigma + m_{j}| = 
|\Lambda|^{N+1}\label{vacpert1}\\
&&\hskip -1cm
\sum_{j=1}^{N+1} \arg {\sigma + m_{j}\over\Lambda} = 2 p\pi\, 
,\quad p\in {\mathbb Z}\, .\label{vacpert2}\\ \nonumber
\end{eqnarray}
When $\Lambda\rightarrow 0$, (\ref{vacpert1}) implies unambiguously 
the classical result (\ref{vaccl}). For $|m_{j}-m_{i}|>>\Lambda$, 
(\ref{vacpert1}) and (\ref{vacpert2}) imply a unique convergent instanton 
series expansion for $\langle\sigma\rangle _{i}$,
\begin{equation}
\label{iexp}
\langle\sigma\rangle _{i} = -m_{i} +\sum_{k=1}^{\infty} 
c_{k}^{(i)}(m_{j})\Lambda ^{k(N+1)}\, ,
\end{equation}
where the $c_{k}^{(i)}$s are calculable functions of the $m_{j}$s, for 
example $c_{1}^{(i)}=1/\prod_{j\not =i}(m_{j}-m_{i})$. The important point 
is that the series (\ref{iexp}) does not depend on the ambiguity in 
(\ref{vacpert2}).

When the vacuum is no longer weakly coupled, instanton calculus is 
plagued by infrared divergences and the semiclassical approximation is 
no longer valid. In particular, the series (\ref{iexp}) has a finite 
radius of convergence. In the strongly coupled regime,
the original field variables strongly fluctuates,
the classical geometric picture of a $\cpN$ target space is lost, and 
the arguments leading to the quantization condition (\ref{instq}) do 
not apply. The $\langle\sigma\rangle_{i}$s can nevertheless be 
calculated, because ${\cal N}=2$ supersymmetry implies that they are 
the analytic continuations of the series (\ref{iexp}). The analytic 
continuations are easily found by noting that those series are the 
$N+1$ solutions of
\begin{equation}
\label{vac}
\prod_{i=1}^{N+1}(\sigma + m_{i}) = \Lambda^{N+1}\, ,
\end{equation}
and that analyticity implies that (\ref{vac}) is always valid. The 
unambiguous vacuum equation (\ref{vac}) is obtained by integrating
\begin{equation}
\label{dwds}
{\partial\weff\over\partial\Sigma} = {1\over 4\pi}\, 
\ln\prod_{i=1}^{N+1} {\Sigma + m_{i}\over\Lambda}\, \cvp
\end{equation}
which yields
\begin{equation}
\label{wex}
\weff (\Sigma,m) = {1\over 4\pi}\, \Sigma 
\ln\prod_{i=1}^{N+1}{\Sigma + m_{i}\over e\Lambda} + {1\over 
4\pi}\sum_{i=1}^{N+1} m_{i}\ln {\Sigma + m_{i}\over e\Lambda}\, \cdotp
\end{equation}
It is interesting to note that this specific formula for $\weff$
corresponds to the lowest energy density at fixed $\sigma$. The 
qualitative difference between the weakly coupled regime where the 
series (\ref{iexp}) converge and the strongly coupled regime where one 
must use the analytic continuations is that in the first instance 
$\langle\sigma\rangle_{i} (\theta) = \langle\sigma\rangle_{i} 
(\theta + 2\pi)$ while in the second instance the different vacua are 
mixed up when $\theta\rightarrow\theta + 2\pi$ \cite{fertheta}. At 
strong coupling, we see that the apparent ambiguity in $\tilde W_{\rm eff}$ 
simply corresponds to a choice of a particular vacuum. The resolution 
of the difficulty associated with the branch cut in the formulas 
(\ref{Wpert}) or (\ref{wex}) is thus qualitatively different at weak 
coupling and at strong coupling (instantons or choice of vacuum), but 
the physics described by (\ref{wex}) is always consistent and 
unambiguous.
\vfill\break

\section{The BPS mass formula at large $N$}
\setcounter{equation}{0}
In this Section, we start the study of the large $N$ limit of our 
model, restricting our attention to the exactly known
central charge (\ref{Zeq}), in strict parallel to what was done in 
four dimensions in \cite{fer1,fer2}. In Section 4, we will show that 
the results obtained by studying this restricted class of observables 
do generalize to the full theory.
\subsection{The enhan\c con and critical points}
To study the large $N$ limit of the central charge, one must first 
study the large $N$ limit of the roots of the equation (\ref{vac}).
This problem was solved in Section 3.2 of \cite{fer1}, and we briefly 
recall the results below. It turns out that
a consistent large $N$ limit is approached when the mass density
\begin{equation}
\label{dens}
\rho_{N}(m) = {1\over N+1}\sum_{i=1}^{N+1}\delta^{(2)}(m-m_{i})
\end{equation}
goes to a well defined distribution $\rho_{\infty}$ when $N\rightarrow
\infty$. This distribution can be the sum of a smooth function and of
$\delta$ function terms. Studying the full $N+1$ 
dimensional space of mass parameters, or, at large $N$, 
the full space of distributions $\rho_{\infty}$, is not very convenient.
It is more instructive to consider one-dimensional sections of the 
full space, parametrized by a global complex mass scale $v$. Given a 
fixed distribution $\rho_{N}(\nu)$ of dimensionless numbers $\nu_{i}$,
the masses are defined to be $m_{i}^{(v)} = v \nu_{i}$. The associated
density is then
\begin{equation}
\label{disv}
\rho_{N}^{(v)}(m) = {1\over v^{2}}\rho_{N}\Bigl( {m\over v}\Bigr).
\end{equation}
By introducing the dimensionless ratio
\begin{equation}
\label{rdef}
r={v\over |\Lambda|}
\end{equation}
and the polynomials
\begin{equation}
\label{pqdef}
p(x) = \prod_{i=1}^{N+1} (x+\nu_{i})\, ,\quad q(x) = p(x) - 
e^{i\theta}/r^{N+1} = \prod_{i=1}^{N+1}(x-x_{i})\, ,
\end{equation}
one can write the central charge (\ref{Zeq}) as
\begin{equation}
\label{beq}
z = Z/v= {i\sqrt{2}\over 4\pi}\,\sum_{i=1}^{N+1}T_{i}w(x_{i}) +
{1\over\sqrt{2}}\sum_{i=1}^{N+1} \nu_{i}S_{i}\, ,
\end{equation}
where the function
\begin{equation}
\label{beq2}
w(x) = -(N+1)x + \sum_{i=1}^{N+1} \nu_{i}\ln (x + \nu_{i})
\end{equation}
is the field-dependent part of the twisted superpotential 
$\tilde W_{\rm eff}/v$ (\ref{wex}) evaluated at one of the vacua.

If $\rho_{\infty}$ is a smooth function, the physics is weakly coupled in
all the vacua when $|r|\rightarrow\infty$. The ``quantum'' roots $x_{i}$ 
of $q$ and the ``classical'' roots $-\nu_{i}$ of $p$ 
then coincide at large $N$, up to 
exponentially suppressed instanton terms. This picture is valid as 
long as $|r|$ is greater than some critical value $r_{\rm c}$. 
When $|r|\leq r_{\rm c}$, the instanton series can diverge,
and the roots $x_{i}$ gradually arrange themselves along an 
inflating curve in the $x$-plane. This curve is a generalization of
the enhan\c con discussed in \cite{pol}. When $r\rightarrow 0$, the 
enhan\c con eventually eats up all the roots, and approaches a circle 
of radius $1/|r|$. If $\rho_{\infty}$ has $\delta$ 
function terms, some of the vacua are always strongly coupled, 
whatever large $|r|$ is. The roots corresponding to such vacua
are arranged on an enhan\c con for all $r$. Finally, let us note that the 
enhan\c con can have several connected components, associated with 
the connected components of the support of $\rho_{\infty}$.

Of crucial importance to us are the critical points that are obtained 
for some special values of the mass parameters. These critical 
points, or singularities, are physically similar to the
Argyres-Douglas points \cite{AD}, which were argued in \cite{fer2} to 
be the four-dimensional generalizations of the Kazakov critical points 
found in zero dimensional matrix models \cite{kaza}. Mathematically, 
both the two dimensional and four dimensional critical points are 
obtained when the discriminant of the polynomial $q$ (\ref{pqdef}) 
vanishes. At large $N$, it was explained in \cite{fer1} that this 
can happen either when a classical root is eaten up by the inflating 
enhan\c con (first class critical point) or when several connected 
components of the enhan\c con collide with each other (second class 
critical point). Let us emphasize that the distinction between first 
class and second class does not arise because 
the low energy physics is different in the two cases---the 
corresponding CFTs are actually the same--- but because the large $N$ 
expansion behaves differently near a first class or a second class 
singularity. In particular, for a given CFT, the first class and 
second class double scaled theories defined in \cite{fer2} are different. 
In Section 4, we will completely characterize those
theories in the two dimensional setting of the present paper.
\subsection{Fractional instantons and IR divergences}
We now give two concrete examples of a first class and a second class 
singularity in our model. The mass densities are chosen to be the same 
as the Higgs vevs densities of the examples studied in \cite{fer1}. 
For that reason, some of the formulas of the Section 4 of \cite{fer1}
can be used here. One minor difference is that $N$ in 
\cite{fer1} must be replaced by $N+1$. 
We have chosen this convention because, in perturbation theory, 
the $N$ of $\suN$ and of $\cpN$ do play the same 
r\^ole, distinguishing between the topology of the dual 
Feynman graphs (see Section 4).
\subsubsection{An example with a first class singularity}
We choose the distribution 
\begin{equation}
\label{dis1st}
\rho_{N}(\nu) = {N\over N+1}\, \delta^{(2)} \bigl(\nu - 1/(N+1)\bigr) + 
{1\over N+1}\,\delta^{(2)}\bigl(\nu - 1/(N+1) +1\bigr)
\end{equation}
and the $\theta$ angle to be $\theta =\pi$.\footnote{The $\theta$ 
dependence could be absorbed in the phase of the parameter $r$. The choice 
$\theta=\pi$ is convenient to compare with the formulas of \cite{fer1}.} 
The first class singularity occurs at the critical parameter
$r_{\rm c}=(N+1)/N^{1-1/(N+1)}$ 
when the two positive real roots $x_{1}$ and $x_{2}$ of the polynomial 
$q$ defined in (\ref{pqdef}) coincide. 
We want to calculate the central charge of the BPS state that becomes 
massless at the singularity.
At large $N$ and $|r|>1$, we have (see equations (45) and (44) of
\cite{fer1} with $N$ replaced by $N+1$)
\begin{equation}
\label{x1N}
x_{1} = {1\over r} - {r + \ln (r-1)\over N r} + {1\over 2N^2 r}\,
\Bigl( \left( \ln (r-1) \right) ^2 - {2 \ln(r-1)\over r-1} + 2r \Bigr)
+ {\cal O}\bigl(1/N^3 \bigr)
\end{equation}
\begin{equation}
\label{x2N}
x_2=1-1/(N+1) - 1/r^{N+1} + {\cal O}\bigl( 1/r^{2(N+1)} \bigr)
= 1-1/N + 1/N^{2} + {\cal O}\bigl(1/N^3 \bigr).
\end{equation}
We then immediately get, by using (\ref{beq}) and (\ref{beq2}),
\begin{equation}
\label{z1st}
{4 i\pi z\over N\sqrt{2}} = -{r-1\over r} + \ln r + 
{(r-1)( \ln(r-1) -1) \over Nr} + 
{\bigl( \ln (r-1) \bigr)^{2}\over 2N^{2}r} + {\cal O}\bigl(1/N^3 \bigr).
\end{equation}
This formula displays all the important qualitative features of the 
large $N$ expansion of ``instanton generated''
BPS observables \cite{fer1}: the expansion parameter is $1/N$;
the expansion breaks down at the critical point $r=1$; each order in $1/N$ 
is given by a mixing between $\ln r$ terms coming from perturbation 
theory and series in $1/r$ obtained by writing $\ln (r-1) = \ln r + 
\ln (1-1/r) = \ln r -\sum_{k=1}^{\infty} 1/(kr^{k})$. These series in 
$1/r$ are naturally interpreted as coming from fractional instantons 
of topological charge $1/(N+1)$. These fractional instantons would be 
the remnant of the disintegration of large instantons at strong 
coupling \cite{fer1}. Let us emphasize, however, that the fractional 
instanton picture remains elusive, because we have not found the 
corresponding field configurations (that must be singular in the 
original field variables), and also because at large $N$ the
topological charge is vanishingly small.
\subsubsection{An example with a second class singularity}
Let us now suppose that $N+1$ is a multiple of four, choose
$\theta=0$, and consider
\begin{equation}
\label{dis2nd}
\rho_{N}(\nu) = {1\over 2}\, \Bigl( \delta^{(2)} (\nu -1 ) +
\delta^{(2)} (\nu + 1) \Bigr).
\end{equation}
The second class singularity occurs when $r=1$, at the merging of
the roots
\begin{equation}
\label{xpm}
x_{1}=\sqrt{1-1/r^{2}}
\end{equation}
and $x_{2}=-x_{1}$ of the polynomial $q$. The formula (\ref{xpm}) is 
exact. In particular, the large $N$ expansion of the roots near a 
second class singularity, though non-analytic, is not blowing up. 
Unlike the four dimensional case, this implies that the large $N$ 
expansion of the central charge itself is not divergent. The exact 
formula is easily derived,
\begin{equation}
\label{z2nd}
{2 i\pi z\over N\sqrt{2}} = \Bigl( 1+{1\over N} \Bigr) \Bigl(
\ln r - \sqrt{1-1/r^{2}} + \ln \left(1+\sqrt{1-1/r^{2}}\right) \Bigr).
\end{equation}
We do not find divergences when $r\rightarrow 1$ because the large $N$ 
expansion of $z$ has only two terms.
We will see in Section 4 that the $1/N$ expansion of more general 
observables has an infinite number of terms and does suffer from IR 
divergences at the critical point. The physical origin of these 
divergences is actually 
the same for first class or second class singularities.

It is interesting to note that we get fractional instanton series 
(presently of topological charge $2/(N+1)$) in the exact formula 
(\ref{z2nd}). The reason is that due to the special choice for the 
density (\ref{dis2nd}), the $1/N$ expansion has only two terms and 
gives the exact answer in this example.
Generically, we expect to get fractional instanton series only at large 
$N$. This simply comes from the fact that, in order to obtain a smooth 
$N\rightarrow\infty$ limit, one must take the $1/(N+1)$th power
of (\ref{vac}) and write
\begin{equation}
\label{vacN}
\exp\int\! d^{2} m\, \rho_{N}(m)\ln (\sigma + m) = \Lambda\, .
\end{equation}
\subsection{The double scaling limits}
\subsubsection{First class}
Let us consider the scaling
\begin{equation}
\label{sca1}
N\rightarrow\infty\, ,\quad \delta = r-1 \rightarrow 0\, ,\quad 
N\delta -\ln N = {\rm constant} = 1/\kappa + \ln\kappa\, ,
\end{equation}
which was used in similar circumstances in four dimensions 
(equations (52) and (54) of \cite{fer2}). From (\ref{z1st}), one can 
easily show that subtle cancellations make the first three terms in the 
perturbative expansion of the amplitude
\begin{equation}
\label{amp1def}
{\cal A}={4i\pi Nz\over\sqrt{2}}
\end{equation}
finite in the double scaling limit (\ref{sca1}),
\begin{equation}
\label{amp1pert}
{\cal A}_{\rm scaled} =
{1\over 2\kappa^{2}} - {1\over\kappa} + {\cal O}(\kappa)\, .
\end{equation}
Going beyond the perturbative expansion is actually easy.
The exact formula for $\cal A$ before the scaling is
\begin{equation}
{\cal A} =N \int_{x_{2}}^{x_{1}}\! dx\, \ln\left[ -\Bigl( x+{1\over 
N+1} \Bigr)^{N} \Bigl(x+{1\over N+1} -1 \Bigr) r^{N+1} \right] .
\end{equation}
By changing the variable to $u=N - (N+1)x$, we immediately see that 
in the limit (\ref{sca1}) we have
\begin{eqnarray}
\label{amp1sca}
{\cal A}\rightarrow 
{\cal A}_{\rm scaled} &=& \int_{1/\kappa}^{\xi (e^{-1/\kappa}/\kappa )}\!
du\,\ln {e^{u-1/\kappa}\over u\kappa}\nonumber\\
&=& \left[ {u^{2}\over 2} - {u\over\kappa} - u\ln {\kappa u\over e}
\right]_{1/\kappa}^{\xi (e^{-1/\kappa}/\kappa )},\\ \nonumber
\end{eqnarray}
where the function $\xi$ is defined by
\begin{equation}
\label{zetadef}
v=ue^{-u} \Longleftrightarrow u=\xi (v)\quad {\rm for}\ u\in [0,1] .
\end{equation}
This explicitly demonstrate that the suitably rescaled central charge 
(\ref{amp1def}) has a finite limit in the scaling (\ref{sca1}). The 
function $\xi$ gives purely non-perturbative contributions 
proportional to $e^{-1/\kappa}$, and thus the perturbative expansion 
is entirely given by the first two terms (\ref{amp1pert}).

One might wonder to what extent the result (\ref{amp1sca}) is 
universal, and whether one can find generalizations. A straightforward 
argument for universality, which was given in \cite{fer2}, is that the 
scaling (\ref{sca1}) corresponds to a low energy limit. Indeed, the 
central charge is related to a mass scale by (\ref{MBPS}), and 
equation (\ref{amp1def}) shows that the 
amplitude having a finite limit in the scaling is $N$ times this mass 
scale. This implies that only the light degrees of 
freedom, that become massless at the singularity, 
survive in the scaling limit. The result (\ref{amp1sca}) suggests that 
the double scaled theory, which must describe the interactions between 
those light degrees of freedom, is a field theory with 
an effective superpotential
\begin{equation}
\label{scaeff}
w_{\rm eff}(u) = {u^{2}\over 2} - {u\over\kappa} - u\ln {\kappa u\over e}
\, \cdotp
\end{equation}
We will be able to characterize this theory in Section 4, and 
universality will then be obvious. Right now, we can discuss
the dependence of the superpotential (\ref{scaeff}) on the particular
choice (\ref{dis1st}). It is actually not difficult to treat the 
general case where $m$ classical roots, for example $x=-\nu_{1},\ldots
x=-\nu_{m}$, melt into the enhan\c con. The starting point is the 
formula
\begin{equation}
\label{genamp}
{\cal A} = N\int\! dx\, \ln \Bigl[ e^{-i\theta}r^{N+1}\prod_{i=1}^{N+1} 
(x+\nu_{i}) \Bigr]
\end{equation}
where the upper and lower bounds of the integral are two distinct 
zeros of the integrand. The density for the roots on the enhan\c con
\begin{equation}
\label{disar}
d_{N}(\nu) = {1\over N+1-m}\sum_{i=m+1}^{N+1}\delta^{(2)}(\nu-\nu_{i})
\end{equation}
is taken to be arbitrary as long as it goes to a well-defined 
distribution $d(\nu)$ of bounded support when $N\rightarrow\infty$. 
We want to see to what extent the double scaled amplitude depends on
$d(\nu)$. The critical point occurs when the classical roots 
$\nu_{i}$, $1\leq i\leq m$, melt into the enhan\c con. As we will see, 
by adjusting the critical value of $r$, we can assume that the 
critical value of the
$\nu_{i}$s for $1\leq i\leq m$ is an arbitrary number $M$ 
such that $|-M+\nu|>\epsilon>0$ for all $\nu$ in the support of 
$d(\nu)$. Let us thus define 
$x=-M + v/N$ and $\nu_{i} = M + v_{i}/N$ for $1\leq i\leq m$. We can write
\begin{eqnarray}
\prod_{i=1}^{N+1}\!\!\left[r(x+\nu_{i})\right] \!\!\!\!&=&\!\!\!\!
\prod_{i=1}^{m} (v+v_{i}) \exp\Bigl[
\ln (r^{N+1} N^{-m}) + \!\!\!\sum_{i=m+1}^{N+1}\!\!\ln (-M+\nu_{i}) +
{v\over N}\sum_{i=1}^{N+1} {1\over -M+\nu_{i}} \Bigr]\nonumber\\
&=&\!\!\!\! \prod_{i=1}^{m} (v+v_{i}) \exp\Bigl[ N(A+\ln r) - m\ln N
+ \ln r + B + C v \Bigr] ,\label{expan}\\ \nonumber
\end{eqnarray}
where we have neglected terms that will go to zero when $N\rightarrow\infty$, 
and $A$, $B$ and $C$ are some constants depending on the 
distribution (\ref{disar}) but not on $N$. In particular, the critical 
value of $r$ is
\begin{equation}
\label{rcgen}
r_{\rm c} = e^{-A} = \exp\Bigl[ -\int\! d^{2}\nu\, d(\nu) \ln (-M+\nu) 
\Bigr]\, .
\end{equation}
The generalized double scaling limit
\begin{eqnarray}
& N\rightarrow\infty\, ,\quad r\rightarrow e^{-A}\, ,\quad
\nu_{i}\rightarrow -M\, ,\quad N(\nu_{i} + M) = {\rm cst}= v_{i}= 
u_{i}/C\, , &\nonumber\\ &
\quad N (A+\ln r) - m\ln (CN) + B - A - i\theta = 
{\rm cst}=1/\kappa + m\ln\kappa & \label{genscaling}\\ \nonumber
\end{eqnarray}
then yields 
\begin{equation}
\label{amp1scagen}
{\cal A}_{\rm scaled} = {1\over C} \Bigl( w_{\rm eff}(u_{2}) - w_{\rm 
eff}(u_{1}) \Bigr)
\end{equation}
where $u_{1}$ and $u_{2}$ are roots of the equation
\begin{equation}
\label{rg}
e^{u} \prod_{i=1}^{m}(u+u_{i})  = {e^{-1/\kappa}\over \kappa^{m}}
\end{equation}
and
\begin{equation}
\label{weffgensca}
w_{\rm eff}(u) = {u^{2}\over 2} + {u\over\kappa }
 + \sum_{i=1}^{m}(u+u_{i})\ln {\kappa (u+u_{i})
\over e}\, \cdotp
\end{equation}
The parameters $u_{i}$ were chosen without loss of generality
such that $\sum_{i=1}^{m} u_{i} 
=0$. We see that all the dependence in the general density 
(\ref{disar}) we started from is in a trivial global finite 
factor $C$. In particular, the formula (\ref{scaeff}) for the case $m=1$ is
recovered after changing $u$ in $-u$.
\subsubsection{Second class}
The general case of an $m$th order critical point can be described by 
choosing $N+1$ to be a multiple of $m$ and
the polynomial $p$ of equation (\ref{pqdef}) to be \cite{fer2}
\begin{equation}
\label{psca2}
p(x) = \biggl( x^{m} + \sum_{k=2}^{m-1} u_{m-k} x^{m-k} + 1 
\biggr)^{(N+1)/m}.
\end{equation}
Defining
\begin{equation}
\label{sca2}
t_{0} = m\ln (r^{N+1} e^{-i\theta})\, ,\quad t_{j} = 
N^{1-j/m}u_{j}\, ,\quad T(u) = \sum_{k=0}^{m-2} t_{k}u^{k} + u^{m}\, ,
\end{equation}
and taking the $N\rightarrow\infty$ limit keeping fixed the $t_{j}$s, 
we see that the amplitude
\begin{equation}
\label{amp2def}
{\cal A} = {2i\pi m N^{1/m} z\over\sqrt{2} t_{0}^{1/m}} = 
{mN^{1/m}\over 2t_{0}^{1/m}}\int\! dx\, \ln \Bigl[ e^{-i\theta}r^{N+1}p(x)
\Bigr]
\end{equation}
has a finite limit
\begin{equation}
\label{amp2sca}
{\cal A}_{\rm scaled} = {1\over 2t_{0}^{1/m}}\int\! T(u)\, du\, ,
\end{equation}
where the upper and lower bounds of the integral are roots of the 
polynomial $T$. This formula strongly suggests that the double scaled
theory is a simple Landau-Ginzburg field theory with a superpotential
\begin{equation}
\label{weff2}
w_{\rm eff} = \sum_{k=0}^{m-2}{t_{k}u^{k+1}\over k+1} + {u^{m+1}\over 
m+1}\, \cdotp
\end{equation}
For example, if $t_{k}=0$ for $k\geq 1$ and $t_{0}=1/\kappa$ we get 
the roots
\begin{equation}
\label{roots}
u_{j}=t_{0}^{1/m}\exp(i\pi (1+2j)/m)
\end{equation}
and the amplitudes
\begin{equation}
\label{casp}
{\cal A}_{jk} ={\kappa^{1/m}\over 2}\int_{u_{k}}^{u_{j}}\! T(u)\, du =
e^{i\pi (j+k+1)/m}\sin(\pi (j-k)/m)\, {\cal I}_{m}(\kappa)\, ,
\end{equation}
with
\begin{equation}
\label{Iint}
{\cal I}_{m}(\kappa) = {m\over (m+1)\kappa }\,\cdotp
\end{equation}
Equations (\ref{amp2sca}), (\ref{casp}) and (\ref{Iint}) are the 
exact analogues of equations (41), (42) and (43) of \cite{fer2}.

\section{The full large $N$ expansion}
\setcounter{equation}{0}
The results of the previous Section suggest that, if we rescale the 
space-time variables from $x^{\mu}$ to $\sigma^{\mu}$,
\begin{equation}
\label{stsca}
\sigma^{\mu} = N^{-1/p} x^{\mu}\, ,
\end{equation}
and take the double scaling limit (\ref{genscaling}) (for $p=1$) or
(\ref{sca2}) (for $p=m$), then the original non-linear
$\sigma$ model tends to a 
well defined ``double scaled'' theory describing the interactions between
the light degrees of freedom. A full proof of this statement of 
course requires to study the full path integral, not only the central 
charge, and that's precisely what we intend to do in this Section.
However, before entering 
into the details, it is useful to give a qualitative discussion that 
applies to the more difficult case of gauge theories as well.

An important point is that, even 
though the double scaling limits correspond to low energy limits, 
as (\ref{stsca}) clearly shows, {\it the limiting procedure does not 
introduce a cut-off}. This means that the resulting theories must be 
defined on all scales, and are thus fully consistent relativistic 
quantum theories, obtained from an asymptotically free quantum field
theory by taking a consistent limit. This fact is particularly 
startling in four dimensions, where the double scaled theories 
are relativistic quantum theories of {\it light} electrically and 
magnetically charged particles, for which only effective descriptions 
were known.

A very elegant, if only heuristic,
way to elucidate the nature of the double scaled 
theories is to introduce a dual representation for the Feynman graph, 
and realize that very large Feynamn graphs dominate near the critical 
points. This classic analysis \cite{review1,review2}, that we sketch 
in the next subsection, 
suggests that the four dimensional double scaled theories are 
string theories while the two dimensional double scaled theories 
are field theories. We then proceed to an explicit proof of this 
result in two dimensions, where the large Feynman graphs of the 
original non-linear $\sigma$ model can be explicitly summed up.
\subsection{Loops of bubbles and the continuum limit}
A generic observable of the two dimensional $\cpN$ model can be 
expanded at large $N$ as a power series in $1/N$,
\begin{equation}
\label{lNs1}
{\cal A} = N^{\alpha}\sum_{h\geq 0} N^{1-h} A_{h}\, ,
\end{equation}
where $N^{\alpha}$ is some normalization that insures that $\cal A$ 
has a finite limit in the double scaling. The coefficients $A_{h}$
can pick contributions both from Feynman diagrams and from non-perturbative 
effects. In the case of $\suN$ gauge theories, Feynman diagrams generate a 
series in $1/N^{2}$, while non-perturbative effects can contribute at 
all orders in $1/N$ \cite{fer1}. 
In Section 3, we have discussed observables for which perturbation 
theory was trivial. However, many other observables are dominated by 
the Feynman graphs contributions. An example that we will discuss 
explicitly below is the mass of non-BPS states. For those, we can 
write
\begin{equation}
\label{Apert}
A_{h} = \sum_{k\geq 0} A_{h,k} \lambda^{2k}\, ,
\end{equation}
where $\lambda = g^{2}N$ is the renormalized 't Hooft coupling constant.
For the double scaling limits to yield a finite result, it is 
necessary that the 
coefficients $A_{h}(\lambda)$ diverge near the critical 
points $\lambda=\lambda_{\rm c}$, at least for sufficiently large $h$.
The whole idea of the double scaling limits is actually that those
divergences are specific enough so that they can be 
compensated for by taking the $N\rightarrow\infty$ limit together with 
the $\lambda\rightarrow\lambda_{\rm c}$ limit.
Typically, one has, up to logarithmic terms, 
\begin{equation}
\label{asy1}
A_{h} \mathop{\propto}\limits _{\lambda\rightarrow\lambda_{\rm c}} 
{1\over (\lambda - \lambda_{\rm c})^{\gamma_{h}-2}}\, \cvp
\end{equation}
where $\gamma_{h}$ is some susceptibility. This shows that near 
$\lambda=\lambda_{\rm c}$, the terms with a high power of $k$ 
dominate in (\ref{Apert}). Those terms are generically associated with 
very large Feynman diagrams, containing a lot of interaction vertices.

\begin{figure}
\centerline{\epsfig{file=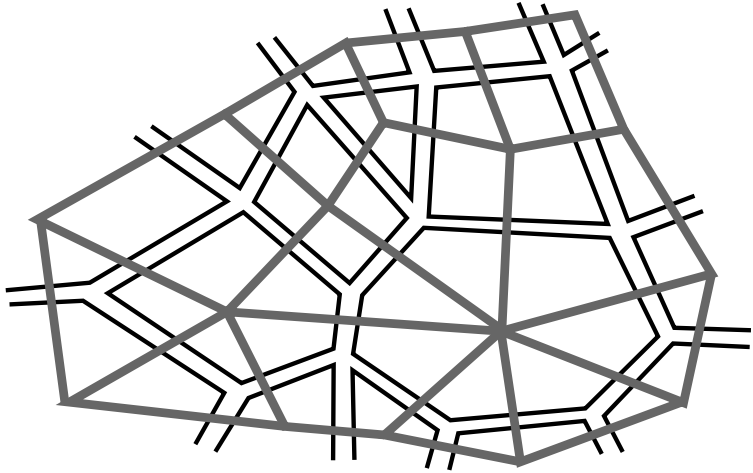,width=10cm}}
\caption{Part of a typical large Feynman diagram in an $\suN$ gauge 
theory. A dual representation (gray lines)
is obtained by associating a $p$-gon to 
each vertex of order $p$. The $p$-gons generate a discretized Riemann 
surface. The double scaling limits correspond to a continuum limit 
where the number of $p$-gons becomes infinite and thus the discretized 
Riemann surfaces become genuine smooth world sheets. The power of 
$\kappa^{2}\propto 1/N^{2}$ counts the genus of these world sheets.
\label{mat}}
\end{figure}
\begin{figure}
\centerline{\epsfig{file=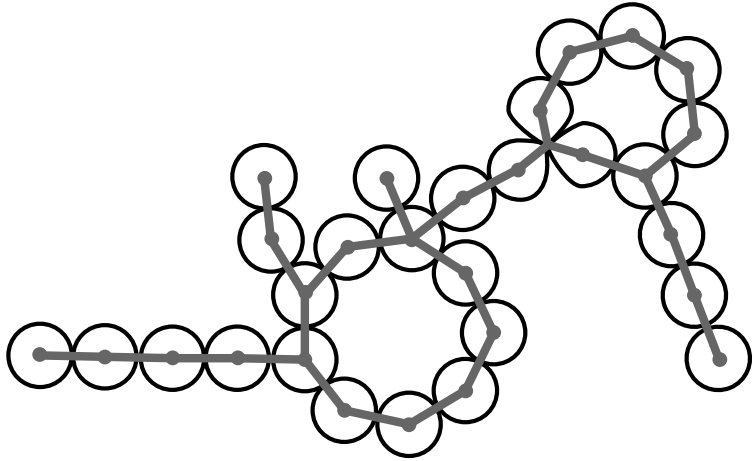,width=10cm}}
\caption{Part of a typical large Feynman diagram in the $\cpN$ 
non-linear $\sigma$ model. A dual representation (gray lines)
is obtained by associating a bound linking $p$ ``molecules'' 
(small gray disks) to each vertex of order 
$2p$. The bounds generate a discretized ``polymer'' with both forced 
(for bounds with $p\geq 3$) and dynamical branching. 
The double scaling limits correspond to a continuum limit 
where the number of bounds become infinite and thus the discretized 
polymers become genuine smooth world lines. The power of 
$\kappa\propto 1/N$ counts the number of loops (two in our example)
of these world lines.
\label{poly}}
\end{figure}

What do those diagrams look like? In the case of gauge theories 
(Figure 1), the answer \cite{tHooft} is that the diagrams contributing
to a given order in $1/N^{2}$ can be mapped to discretized Riemann 
surfaces of a given genus. Large diagrams have 
a very large number of polygons, and thus the double scaling limit is 
a continuum limit for the discretized surfaces. We conclude that the 
resulting theory must be a string theory. For
non-linear $\sigma$ models the analogous statements are 
easy to derive. In the case of linear $\sigma$ models \cite{review1}, 
the typical large $N$ graphs are ``bubble'' diagrams, the order in 
$1/N$ being related to the number of loops of bubbles. In a dual 
representation (Figure 2), we obtain a discretized world line (or 
``polymer'') with a given number of loops. The double scaling limit 
is then a continuum limit for these discretized loop diagrams, and 
as a result we should obtain a standard field theory. The case of 
non-linear $\sigma$ models is similar, with the additional subtlety 
that we have interaction vertices of any order $p\geq 2$, but we still 
expect the resulting theory to be a field theory.

For the purposes of the calculations that follow, it is convenient to 
go to the euclidean for which 
$x^{2}=ix^{0}$ and $\partialsl _{\rm Eucl.} = -i \partialsl _{\rm Mink.}$.
The euclidean lagrangian deduced from (\ref{lag}) is
\begin{eqnarray}
\label{lagE}
&&\hskip -1.8cm L_{\rm E}=-L=\sum_{i=1}^{N+1}\Bigl( 
\nabla_{\alpha}\phi_{i}^{\dagger}\, \nabla_{\alpha}
\phi_{i} + |\sigma + m_{i}|^{2}\, |\phi_{i}|^{2} +\bar\psi_{i}\left(
\nablasl - \check\sigma - \check m \right)\psi_{i}\Bigr)
+{i\theta\over 2\pi}\, *v\nonumber\\
&&\hskip 1.5cm  
- D\,\Bigl( \sum_{i=1}^{N+1}|\phi_{i}|^{2} - {4\pi\over g^{2}}\Bigr) - 
i\sqrt{2}\,\bar\mu\sum_{i=1}^{N+1} \psi_{i}\phi_{i}^{\dagger} +
i\sqrt{2}\sum_{i=1}^{N+1}\phi_{i}\bar\psi_{i}\,\mu\, ,\\ \nonumber
\end{eqnarray}
where we have defined
the covariant derivative $\nabla_{\alpha} = \partial_{\alpha} + 
iv_{\alpha}$, the field strenght $*v=\partial_{1}v_{2}-\partial_{2}v_{1}$, 
and $\check\sigma = \re\sigma - i\sigma^{3}\im\sigma$, $\check m_{i} =  
\re m_{i} - i\sigma^{3}\im m_{i}$.
\subsection{First class singularities}
We consider the distribution (\ref{dis1st}) again. It is convenient to 
make the substitution $\sigma\rightarrow \sigma - v/(N+1)$ and to define
\begin{equation}
\label{rdef1}
m=-v=\Lambda r,
\end{equation}
in line with the notations of Section 3.
The large $N$ limit is studied by integrating the superfields 
$\Phi_{i}=(\phi_{i},\psi_{i})$ from 
(\ref{lagE}). This can be done exactly, and yields a non-local effective 
action proportional to $N$. The $1/N$ expansion is then a perturbative 
expansion for this non-local effective action. For the particular 
distribution (\ref{dis1st}), the superfield $\Phi_{N+1}$ plays a special 
r\^ole. One must actually keep this superfield explicitly in order to get a 
well-defined saddle point at large $N$. We are then left with a path 
integral over the fields $\phi = \phi_{N+1}/\sqrt{N}$, $\psi = 
\psi_{N+1}/\sqrt{N}$ and $\Sigma = (\sigma,\mu,D,v)$ which reads
\begin{eqnarray}
\label{pi1st}
&&\hskip -1cm \int\! {\cal D}(\phi,\psi,\sigma,\mu,D,v_{\alpha})\,
\left[ {\det_{\rm r}(\nablasl - \check\sigma)\over\det_{\rm 
r}(-\nabla^{2} + |\sigma|^{2} - D - 2\, \bar\mu (\nablasl 
-\check\sigma )^{-1}\mu )}\right]^{N} \nonumber\\
&& \exp\Biggl[ -N \int\! d^{2}x\,\Bigl( 
\nabla_{\alpha}\phi^{\dagger}\, \nabla_{\alpha}
\phi + (|\sigma + m|^{2}-D)\, |\phi|^{2} +\bar\psi\left(
\nablasl - \check\sigma - \check m \right)\psi\nonumber\\
&&\hskip 3.7cm  +{i\theta\over 2\pi N}\, *v
+ {D\over 2\pi} \ln {\mu\over |\Lambda|} - 
i\sqrt{2}\,\bar\mu\psi\phi^{\dagger} +
i\sqrt{2}\phi\bar\psi\,\mu\, \Bigr)\Biggr] . \label{detrap} \\\nonumber
\end{eqnarray}
The renormalized determinants are studied in Appendix A, where all the 
formulas that we will use in the following can be found. The scale $\mu$ 
appearing in $\ln (\mu/|\Lambda|)$, not to be confused with the spinor 
$\mu$, is a renormalization scale appearing in the definition of the 
determinants.
\subsubsection{The BPS/anti-BPS bound state}
Before taking the scaling limit (\ref{sca1}) on the 
full path integral (\ref{pi1st}), it is instructive to study explicitly
an observable that has a non-trivial perturbative expansion, to 
complement the discussion of Section 3. We 
will consider the mass of a BPS/anti-BPS bound state that turns out to 
become massless at the first class singularity.

The saddle point equations for (\ref{pi1st}) are deduced from the effective 
potential\footnote{We have chosen the background electric field to be zero. 
A possible non-zero electric field is an effect of order 
$1/N$, and thus can be neglected in the saddle point equations.}
\begin{equation}
\label{pot1}
V_{\rm eff} = (|\sigma + m|^{2} - D)\, |\phi|^{2} + {D\over 2\pi}
\ln {\mu\over |\Lambda|} + {|\sigma|^{2}\over 4\pi} \ln {|\sigma|^{2}\over 
e\mu^{2}} - {|\sigma|^{2}-D\over 4\pi} \ln {|\sigma|^{2}-D \over 
e\mu^{2}}\, \cdotp
\end{equation}
The saddle points $dV_{\rm eff}=0$ 
correspond to the possible $N=\infty$ limit of the vacuum expectation
values of the fields, and are also given by the condition 
$V_{\rm eff} = 0$ by supersymmetry. 
Two cases must be considered. When $|r|=|m/\Lambda| <1$, the only 
solution is
\begin{equation}
\label{spsc1}
\langle\phi\rangle = \langle D\rangle = 0\, ,\quad 
|\langle\sigma\rangle | = |\Lambda|\, .
\end{equation}
We thus get an enhan\c con, as discussed in Section 3.1 or
in more details in \cite{fer1}, which is simply a circle in the $\sigma$ 
plane. When $|r|>1$, in addition to the enhan\c con (\ref{spsc1}), 
we get another solution,
\begin{equation}
\label{spwc1}
\langle D\rangle = 0\, , \quad |\langle\phi\rangle |^{2}= {1\over 2\pi}\ln 
{|m|\over |\Lambda|}\, \cvp\quad\langle\sigma\rangle = -m\, .
\end{equation}
This solution corresponds to the root $x_{2}$ in equation (\ref{x2N}),
while the root $x_{1}$ of equation (\ref{x1N}) lies on the 
enhan\c con (\ref{spsc1}). The critical 
point occurs when the two roots coincide.

The vacuum (\ref{spwc1}) is weakly coupled when $|m|\gg |\Lambda|$. In 
this regime, the relevant fields are the coordinates  
$\phi_{1}/\phi_{N+1},\ldots ,\phi_{N}/\phi_{N+1}$ on $\cpN$. They create BPS 
states of mass $|m|$. The field $\phi = \phi_{N+1}/\sqrt{N}$ plays the 
r\^ole of a Higgs field breaking the ${\rm U}(1)$ gauge symmetry 
(\ref{cpNid}). Choosing the unitary gauge and using (\ref{spwc1}), we 
write
\begin{equation}
\label{higgs}
\phi = \left( {1\over 2\pi}\ln {|m|\over |\Lambda|}\right) ^{1/2} \!\! + 
\varphi = {2\sqrt{\pi}\over g(|m|)} + \varphi\, \cvp
\end{equation}
where $\varphi$ is a fluctuating real scalar field. 
The constraint (\ref{cpNeq}), which is valid in the vacuum 
(\ref{spwc1}), shows that $\varphi$ is a composite 
operator creating a two-particle BPS/anti-BPS bound state of the 
elementary quanta. Though the attractive force between these quanta is 
of order $1/N$ at large $N$, the mixing between the $N$ flavors will 
stabilize the bound state significantly, and the binding energy 
should be of order $N^{0}$. The mass $m_{\rm b}$ of this bound state 
is a nice example of an observable which has a highly non-trivial 
perturbative expansion, as opposed to the cases studied in Section 3. 
We can straightforwardly calculate the leading large $N$ approximation 
for $m_{\rm b}$, by looking at the quadratic piece of the effective 
action deduced from (\ref{pi1st}), see Appendix A.  
There is a mixing between $\varphi$ and the gauge multiplet.\footnote{A 
na\"\i ve application of the standard results about the super-Higgs
mechanism in four dimensions suggests that the Higgs and the gauge 
fields actually belong to the same supersymmetry multiplet and have 
the same mass. This is not correct in two dimensions, 
because a non-linear twisted superpotential is generated.} By inverting the 
matrix-valued propagator, we find a pole at $p^{2}=-m_{\rm b}^{2}$ such 
that
\begin{equation}
\label{mb}
\ln \left| {m\over \Lambda}\right| = 
{1+\sqrt{1+u^{2}}\over u} \arctan {1\over u} = 
{8\pi^{2}\over\lambda_{\mu}^{2}} 
-{8\pi^{2}\over\lambda_{\mu,\rm c}^{2}}\, \cvp
\end{equation}
where
\begin{equation}
\label{udef}
u = \sqrt{{4|m|^{2}\over m_{\rm b}^{2}} -1}\,\, ,\quad
{8\pi^{2}\over\lambda_{\mu}^{2}} = \ln {\mu\over |\Lambda|}\, \cvp\quad
{8\pi^{2}\over\lambda_{\mu,\rm c}^{2}} = \ln {\mu\over |m|}\, \cdotp
\end{equation}
If we define perturbation theory in terms of the 't Hooft coupling 
constant $\lambda^{2}_{\mu}$ with renormalization scale $\mu\geq |\Lambda|$,
(\ref{mb}) implies an expansion
\begin{equation}
\label{pexp}
u = {\lambda_{\mu}^{2}\over 8\pi} + \sum_{k=2}^{\infty} 
u_{\mu,k}\lambda^{2k}_{\mu}
\end{equation}
with some $\mu$-dependent coefficients $u_{k}$.
Near the critical point $m\rightarrow |\Lambda |$ or $\lambda_{\mu ,\rm 
c}\rightarrow \lambda_{\mu}$, (\ref{mb}) implies 
\begin{equation}
\label{ucrit}
u\sim {\lambda_{\mu ,\rm c}^{2}/ (8\pi^{2}) \over 1- 
\lambda_{\mu}^{2}/\lambda_{\mu ,\rm c}^{2}}\, \cdotp
\end{equation}
For the general picture of Section 4.1 to apply, one would need to prove 
that the series (\ref{pexp}) has a radius of convergence
\begin{equation}
\label{R}
R_{\mu} = \lambda_{\mu,\rm c}\, ,
\end{equation}
or equivalently that
\begin{equation}
\label{ukas}
u_{\mu,k}\mathop{\sim}\limits _{k\rightarrow\infty}  ={\lambda_{\mu ,\rm 
c}^{2(1-k)}\over 8\pi^{2}}\, \cvp
\end{equation}
at least for a particular choice of $\mu$.
This would indeed imply that the perturbative series (\ref{pexp}) 
is dominated by the terms with large $k$, or equivalently by the large 
Feynman graphs, near the singularity. Proving (\ref{R}), however, turns 
out to be particularly tricky. One can show rigorously that 
$R_{\mu}>0$ for all $\mu$, and that
if $R_{\mu} = \lambda_{\mu,\rm c}$ for some particular $\mu$, then it is 
true for all larger values of $\mu$ as well. One can also show, using 
Picard theorem, that for 
$\mu = |\Lambda|$, which means $\lambda_{\mu,\rm c} = \infty$, the radius 
is actually finite, contradicting (\ref{R}). To really understand what was 
going on, we performed a numerical analysis. For large 
enough values of $\mu$, such that $\lambda^{2}_{\mu,\rm c}/(8\pi^{2})
\lesssim 0.45$, (\ref{ukas}) is found to be satisfied, 
with a rapid and smooth convergence. For 
$\lambda^{2}_{\mu,\rm c}/(8\pi^{2}) \gtrsim 0.45$, however, the
behaviour of the coefficients $u_{\mu,k}$ changes drastically
and (\ref{ukas}) is apparently violated.

The perturbative series for the mass of the bound state itself
can be immediately deduced from (\ref{pexp}) and (\ref{udef}), 
and it has the same properties. At small coupling, we have 
\begin{equation}
\label{mbexp}
m_{\rm b} \simeq 2 |m| \Bigl( 1 - {\lambda_{\mu}^{2}\over 2^{7}\pi^{2}} + 
{\cal O}(\lambda_{\mu}^{4}) \Bigr)\, ,
\end{equation}
but near the critical points the high orders in perturbation theory 
dominate and we find
\begin{equation}
\label{mbcrit}
m_{\rm b} \propto {1\over u} \rightarrow 0\, .
\end{equation}
It is important to realize that the asymptotic behaviours (\ref{ucrit}) or 
(\ref{mbcrit}) do not depend on $\mu$, an obvious consequence of the 
renormalization group equations. If we introduce $\delta = r-1 \sim \ln 
(m/\Lambda)$ as in (\ref{sca1}), the asymptotics read
\begin{equation}
\label{asdelta}
u \sim 1/|\delta|\, ,\quad m_{\rm b} \sim 2 |\Lambda\delta |\, .
\end{equation}
This equation has two important consequences. First it shows that the 
BPS/anti-BPS bound state becomes massless at the critical point in the 
leading large $N$ approximation. This is a nice result, because usually only 
BPS states can be proven to become massless at such singularities, by 
using exact BPS mass formulas as the one discussed in Section 2 or 3
(in our example the massless BPS states are the solitons interpolating 
between the two vacua that ``collide'' at the critical point). Second, 
after doing the rescaling (\ref{stsca}) for $p=1$ (the correct value for a 
first class singularity), the mass $m_{\rm b}$ is multiplied by $N$ (in 
the same way as the central charge was mutiplied my $N$, see 
(\ref{amp1def})), and thus has a finite non-trivial
limit in the double scaling (\ref{sca1}) at leading order,
\begin{equation}
\label{losca1}
(Nm_{\rm b})_{\rm scaled} = 2|\Lambda|/\kappa + {\cal O}(\kappa^{0})\, .
\end{equation}

What about the higher orders in $1/N$? Can we trust the results obtained 
in the leading $1/N$ approximation? As we will show shortly,
qualitatively, the answer is yes: the 
BPS/anti-BPS bound state does become massless, and $Nm_{\rm b}$ does have a 
non-trivial
finite limit in the scaling (\ref{sca1}). However, quantitatively, there 
are some important subtleties. The fact that $(Nm_{\rm b})_{\rm scaled}$ 
is finite to all orders and has a non-trivial expansion in $\kappa$
(a result we will prove in the next subsection) 
implies that the $1/N$ corrections to $m_{\rm b}$ must {\it diverge} at 
fixed $N$ when 
$\delta\rightarrow 0$. The leading order equation (\ref{asdelta}) is thus
not to be trusted. The correct asymptotics is actually
\begin{equation}
\label{ascorr}
m_{\rm b} \propto |\Lambda \delta^{3/2}|\, .
\end{equation}
One must not be confused and conclude that, in the 
exact theory, (\ref{losca1}) is wrong. Equation 
(\ref{ascorr}) is valid when $\delta\rightarrow 0$ at {\it fixed} $N$, 
while in the double scaling limit (\ref{sca1}) we take $N\rightarrow\infty$
and $\delta\rightarrow 0$ in a correlated way. The result 
(\ref{ascorr}) and the fact that $(Nm_{\rm b})_{\rm scaled}$ has a 
non-trivial expansion in $\kappa$, far from being 
contradictory, actually complement each other. The non-trivial exponent 
in (\ref{ascorr}) is a consequence of the fact that the CFT at the 
critical point is non-trivial. This non-triviality is the cause of the 
divergences in the $1/N$ expansion. 
$(Nm_{\rm b})_{\rm scaled}$ in turn picks up the most IR divergent 
terms in this expansion (see also \cite{fer1} for further discussion).
\subsubsection{The double scaling limit}
Showing that the scaling (\ref{sca1}) is fully consistent
might look like a very difficult task, because it amounts to 
resumming the most divergent terms in the $1/N$ expansion to all orders and 
beyond. What makes it possible, and even easy, is the IR nature of the 
divergences. Not surprisingly, and as equation (\ref{stsca}) shows, 
this implies that
the double scaling limit is also a low energy limit, and the path integral 
(\ref{pi1st}) simplifies considerably in such a limit. The same property 
makes tractable the case of linear $\sigma$ models \cite{zi}, and
was first used in the context of non-linear $\sigma$ models in \cite{fer3}.

The starting point of the proof is the non-local effective action defining 
the path integral (\ref{pi1st}). By rescaling $\phi \rightarrow 
\phi/\sqrt{N}$ and introducing the field $s = \sigma + m$ and the 
functionals $\xi$ and $\zeta$ discussed in Appendix A, it reads
\begin{eqnarray}
&&\hskip -1cm  S_{\rm eff} = \int\! d^{2}x\, \Bigl( 
\nabla_{\alpha}\phi^{\dagger}\, \nabla_{\alpha}
\phi + (|s|^{2}-D)\, |\phi|^{2} +{i\theta\over 2\pi}\, *v 
 + {N D\over 2\pi} \ln {\mu\over |\Lambda|}\Bigr) \nonumber\\ 
&&\hskip 1.5cm  +2 N\, \xi\left[ |s-m|^{2} - D,v_{\alpha}\right] - 2 N\,
\zeta\left[ -m +s,v_{\alpha}\strut\right] + {\rm fermions}\, . 
\label{seff1st1} \\ \nonumber
\end{eqnarray}
We will work thereafter with the bosonic fields only, the fermionic part 
of the action being unambiguously determined by supersymmetry. The 
rescaling of the space-time variables (\ref{stsca}) $x^{\alpha} = N\sigma 
^{\alpha}$ implies that a quantity
of dimension $D$ scale as $N^{-D}$. This means that the volume element 
$d^{2}x$ scales as $N^{2}$, the partial derivatives $\partial_{\alpha}$ and
the fields $s$ and $v_{\alpha}$ scale as $1/N$, and the fields $D$ and 
$v_{\alpha\beta}$ scale as $1/N^{2}$. Moreover, (\ref{sca1}) shows that
$\delta$ scales like $1/N$. With those scalings, only a few 
terms in $S_{\rm eff}$ survive when $N\rightarrow\infty$. It is 
straightforward to check that those terms are at most linear in $D$ 
and $*v$, and at most cubic in $s$. 
Terms containing derivatives cannot survive, because Lorentz 
invariance implies that derivatives must come in pair, and the dominant
term with derivatives, $\int (\partial s)^{2}$, goes like $N^{-1}$. 
These remarks imply that all the relevant terms in $S_{\rm eff}$ can be 
obtained from the potential (\ref{pot1}) and from (\ref{linv}). Adding up 
all the contributions, and using $\arg m = \arg\Lambda + \im\delta $, we get
\begin{eqnarray}
&& \hskip -1.5cm S_{\rm eff} = \int_{|\Lambda|}\! d^{2}x\, \Bigl( 
\nabla_{\alpha}\phi^{\dagger}\, \nabla_{\alpha}
\phi + (|s|^{2}-D)\, |\phi|^{2} + {ND\over 2\pi} \re (\delta - s/\Lambda )
\nonumber \\ 
&&\hskip 3cm - {iN *v\over 2\pi} \im (\delta - s/\Lambda )  + {\rm 
fermions} \Bigr) + {\cal O}(1/N). \label{seff1st2} \\ \nonumber
\end{eqnarray}
This formula is strictly valid only with a cut-off $\sim N^{0}|\Lambda|$ 
that we have indicated on the integral sign, 
since we have been using a derivative expansion. 
The terms that potentially scale as $N^{2}$ or $N$ cancel, which is a 
necessary condition for the scaling (\ref{sca1}) to be consistent. The 
terms cubic in $s$ also cancel, as a consequence of supersymmetry.
Back to Minkowski space-time, and adding the fermions,
(\ref{seff1st2}) can be written as
\begin{equation}
\label{seff1st3}
S_{\rm eff} =\int_{|\Lambda|}\! d^{2}x\, \left[ {1\over 4}\int\! d^{4}\theta \,
\Phi^{\dagger} e^{2V}\Phi - {N\over 4\pi} \re\int\! d\tm d\tbp 
\Bigl( \delta\Sigma - {\Sigma^{2}\over 2\Lambda}\Bigr) \right] + {\cal 
O}(1/N).
\end{equation}
It is useful at this point to introduce explicitly the scalings in 
$N$. This can be done in a manifestly supersymmetric way by defining
\begin{equation}
\label{scasusy}
x^{\mu} = N \sigma^{\mu}\, ,\quad \theta_{\pm} = \sqrt{N}\,\Theta_{\pm}\, 
,\quad \bar\theta_{\pm} = \sqrt{N}\,\bar\Theta_{\pm}\, ,
\end{equation}
which yield a new super field strenght
\begin{equation}
\label{scasusy2}
S = \bar D_{+,(\Theta ,\sigma )} D_{-,(\Theta ,\sigma )} V
=N \bar D_{+,(\theta ,x)} D_{-,(\theta ,x)} V  =N \Sigma \,
\end{equation}
and an action
\begin{equation}
\label{seff1st4}
\hskip -.15cm S_{\rm eff} = \!\int_{N|\Lambda|}\!\!\!\! d^{2}\sigma
\left[ {1\over 4}\int\! d^{4}\Theta \,
\Phi^{\dagger} e^{2V}\Phi - {1\over 4\pi} \re\int\! d\Theta_{-}
d\bar\Theta_{+} 
\Bigl( N\delta\, S - {S^{2}\over 2\Lambda}\Bigr) \right] +{\cal O}(1/N).
\end{equation}
The cut-off in the new space-time coordinates $\sigma$ is now of order 
$N$. Neglected terms, that all go
to zero when $N\rightarrow\infty$, include for example the gauge and $s$ 
fields kinetic terms that may be deduced from (\ref{seff1st1}),
\begin{equation}
\label{gfterm}
-{N\over 32\pi |\Lambda|^{2}}\int\! d^{2}x d^{4}\theta\, \bar\Sigma 
\Sigma = -{1\over 32\pi |\Lambda|^{2} N} \int\! d^{2}\sigma 
d^{4}\Theta\, \bar S S\, .
\end{equation}

Let us now actually take the limit $N\rightarrow\infty$ and 
$\delta\rightarrow 0$ in (\ref{seff1st4}). Since the cut-off goes to 
infinity in this limit, one must renormalize the theory 
(\ref{seff1st4}) in order to get a finite answer. This is the origin 
of the logarithmic correction to the na\"\i ve scaling in (\ref{sca1}). 
Only a one-loop renormalization of the linear term in the twisted
superpotential is needed, as can be 
checked by integrating out the superfield $\Phi$. One must add a 
counterterm $S(\ln (\Lambda_{0}/\mu))/(4\pi)$ to make the theory 
finite, where $\Lambda_{0} = N|\Lambda|$ is the cut-off for 
(\ref{seff1st4}) and $\mu$ an arbitrary renormalization scale.
This means that $N\delta$ is renormalized, with
\begin{equation}
\label{Ndren}
N\delta = (N\delta)_{\rm r} + \ln {\Lambda_{0}\over \mu} = 
(N\delta)_{\rm r} +\ln {|\Lambda|\over\mu} + \ln N\, ,
\end{equation}
where $(N\delta)_{\rm r}=1/\kappa + \ln\kappa$ is the renormalized quantity
to be held fixed when the cut-off is removed. We thus recover the scaling
(\ref{sca1}), and this completes the proof.

Several comments are here in order. First, it is important to 
understand the meaning of the ``truncated'' action (\ref{seff1st4}), 
with respect to the full action (\ref{seff1st1}). To do the ordinary 
$1/N$ expansion, one starts from (\ref{seff1st1}) and expands around a 
saddle point, for example (\ref{spsc1}). An infinite number of 
vertices is then generated from (\ref{seff1st1}), a vertex of order 
$p$ contributing with a power of $N^{1-p/2}$ by the standard large $N$ 
counting. The few terms that we have kept in (\ref{seff1st4}) or 
(\ref{seff1st2}), like the terms $|s|^{2}|\phi|^{2}$ or 
$-D|\phi|^{2}$, {\it correspond to the vertices producing the most IR 
divergent contributions near the critical points}, which are the only one 
that survive in the scaling (\ref{sca1}). A second important comment is 
that the double scaled theory does not depend on a cut-off. It is a field 
theory consistent on all scales, defined by the action
\begin{equation}
\label{Sdsca1}
S_{\rm scaled} = \!\int\! d^{2}\sigma
\left[ {1\over 4}\int\! d^{4}\Theta \,
\Phi^{\dagger} e^{2V}\Phi - {1\over 4\pi} \re\int\! d\Theta_{-}
d\bar\Theta_{+} 
\biggl( \Bigl({1\over\kappa} + \ln\kappa + \ln {\Lambda_{0}\over\mu }
\Bigr)S - {S^{2}\over 2\Lambda}\biggr) \right],
\end{equation}
with the UV cut-off $\Lambda_{0}$ taken to infinity and renormalized 
coupling constant $\kappa$. Interestingly, the phenomenon of 
dimensional transmutation takes place, and the coupling $\kappa$ is 
actually replaced by a scale $M$ is the quantum theory,
\begin{equation}
\label{dimtrans}
{1\over\kappa} + \ln\kappa = \ln {\mu\over M}\,\cdotp
\end{equation}
The physics described by the action (\ref{Sdsca1}) depends on the 
dimensionless ratio
\begin{equation}
\label{defR}
R = M/\Lambda\, .
\end{equation}
By integrating out the superfield 
$\Phi$, one can deduce the effective superpotential $w$ by using
(\ref{wex}),
\begin{equation}
\label{w1stsca}
w(s) = {1\over 4\pi} \Bigl[ s \ln {s\over eM} - {s^{2}\over 
2\Lambda}\Bigr] = {\Lambda\over 4\pi}\Bigl[ u\ln {u\over eR} - 
{u^{2}\over 2}\Bigr] \cvp
\end{equation}
with $s=\Lambda u$. We recover, up to a global factor, the result 
obtained in Section 3, equation (\ref{scaeff}). The double scaled
theory has two vacua, 
obtained by solving $dw/du = 0$. When $R = 1/e$, the two vacua collide 
at $u=1$, and we get a critical point, which is nothing but the original 
critical point used to define the double scaling limit. By expanding 
around $u=1$, $S = \Lambda + T$, and by using the formulas of the Appendix, 
one can deduce the low energy effective action describing (\ref{Sdsca1}) 
near the critical point,
\begin{equation}
\label{lscaeff}
S_{\rm scaled, eff} = -\int_{|\Lambda|}\! d^{2}\sigma
\left[ {1\over 32\pi |\Lambda|^{2}}\int\! d^{4}\Theta \,
T^{\dagger}T - {1\over 4\pi} \re\int\! d\Theta_{-} d\bar\Theta_{+} 
\Bigl( T\ln (eR) + {T^{3}\over 6\Lambda^{2}}\Bigr) \right] .
\end{equation}
Not surprisingly, we obtain a simple Landau-Ginzburg description of 
the $A_{1}$ minimal ${\cal N}=2$ CFT. Note that the double scaled 
theory (\ref{Sdsca1}), however, differ from this simple Landau-Ginzburg 
description at high energies.

Let us emphasize that the same qualitative phenomena are likely to 
occur in the gauge/string theory case at a first class singularity 
\cite{fer2}. In particular, the string coupling should dissappear 
and be replaced by a mass scale.
\subsection{Second class singularities}
For the sake of simplicity and conciseness, we will study the case of the
simplest critical point only, corresponding to $m=2$ in the notations of 
Section 3.3.2. Unlike the 
case of the first class singularity, we can integrate over all of the 
$N+1$ superfields $\Phi_{i}$ in (\ref{lagE}), and we are left with 
the following path integral over $\Sigma$,
\begin{eqnarray}
&&\hskip -.6cm 
\int\! {\cal D}(\sigma,\mu,D,v_{\alpha}) \exp\Biggl[ - (N+1)\int\! 
d^{2}x\, \Bigl( {D\over 2\pi} \ln {\mu\over |\Lambda|} + {i\theta\over 
2\pi (N+1)}\, *v \Bigr)\Biggr]\times\label{det2}\\
&&\hskip -1.3cm
\left[ {\det_{\rm r}(\nablasl - \check\sigma - \check m) \det_{\rm r} 
(\nablasl - \check\sigma + \check m) \over \det_{\rm r} 
(-\nabla^{2} + |\sigma + m|^{2} - D - 2\, \bar\mu F_{+}^{-1}\mu ) 
\det_{\rm r} (-\nabla^{2} + |\sigma - m|^{2} - D - 2\,
\bar\mu F_{-}^{-1}\mu )} \right]^{{\scriptstyle N+1\over\scriptstyle 2}}
\hskip -.7cm \cvp\nonumber\\ \nonumber
\end{eqnarray}
where $F_{\pm} = \nablasl - (\check\sigma \pm \check m)$. The 
effective potential is
\begin{eqnarray}
&&\hskip -1.5cm  V_{\rm eff} =  {D\over 2\pi}\ln {\mu\over |\Lambda|} + 
{|\sigma + m|^{2}\over 8\pi} \ln {|\sigma +m|^{2}\over e\mu^{2}} +
{|\sigma - m|^{2}\over 8\pi} \ln {|\sigma -m|^{2}\over e\mu^{2}}
\nonumber\\&&
-{|\sigma + m|^{2}-D\over 8\pi} \ln {|\sigma +m|^{2}-D \over e\mu^{2}}
-{|\sigma - m|^{2}-D\over 8\pi} \ln {|\sigma -m|^{2}-D \over e\mu^{2}}
\, \cvp\label{pot2nd}\\ \nonumber
\end{eqnarray}
and yields the $N\rightarrow\infty$ vacuum expectation values
\begin{equation}
\label{vac2nd}
\langle D\rangle = 0\, , \quad |\langle\sigma\rangle^{2} - 
m^{2} | = |\Lambda|^{2}\, .
\end{equation}
The solution for $\sigma$ gives the enhan\c con that was described in 
the Figure 4 of \cite{fer1}. The critical point 
is obtained when the two disconnected components of the enhan\c con 
collide, at $|m|=|\Lambda|$ and $\langle\sigma\rangle = 0$.

To study the double scaling limit, we use
the same strategy as in Section 4.2.2. We focus on the bosonic part 
of the action. The rescaling of the 
space-time variables is $x^{\alpha} = \sqrt{N}\sigma^{\alpha}$ 
(\ref{stsca}), showing that $d^{2}x$ scales as $N$, 
the partial derivative $\partial_{\alpha}$ and the fields $\sigma$ 
and $v_{\alpha}$ scale as $1/\sqrt{N}$, and the fields $D$ and 
$v_{\alpha\beta}$ scale as $1/N$. As for the deviation from the critical
point,
\begin{equation}
\label{defdel2nd}
\delta = \ln {m^{2}\over\Lambda^{2}}\, \cvp
\end{equation}
equation (\ref{sca2}) shows that it scales as $1/N$.
Those scalings imply that the only terms 
that can survive are either the kinetic term for $\sigma$ or the 
gauge field, that can be derived from (\ref{les2}) and (\ref{lez2}),
potential terms 
at most quartic in $\sigma$ and quadratic in $D$, that can be derived 
from (\ref{pot2nd}), and terms linear in $v_{\alpha\beta}$ and at 
most quadratic in $\sigma$ that can be derived from (\ref{linv}). 
Adding up all the contributions, we get the action
\begin{eqnarray}
&&\hskip -1.5cm S_{\rm eff} = (N+1) \int_{|\Lambda|}\! d^{2}x\, \left[
{1\over 8\pi |m|^{2}} \Bigl( {1\over 2} 
v_{\alpha\beta}v_{\alpha\beta} + 
\partial_{\alpha}\sigma^{\dagger}\partial_{\alpha}\sigma - 
D^{2}\Bigr)\right. \nonumber\\
&&\hskip -.5cm\left. + {D\over 4\pi} \re ( \delta - \sigma^{2}/m^{2}) - 
{i*v\over 4\pi} \im ( \delta - \sigma^{2}/
m^{2} ) + {\rm fermions}\right] + {\cal O}(1/\sqrt{N}) .\label{seff2nd}\\
\nonumber
\end{eqnarray}
Introducing the scalings explicitly,
\begin{equation}
\label{scasusy3}
x^{\mu}=\sqrt{N}\sigma^{\mu}\, ,\quad \theta_{\pm} = 
N^{1/4}\, \Theta_{\pm}\, ,\quad \bar\theta_{\pm} = 
N^{1/4}\, \bar\Theta_{\pm}\, , \quad S = \sqrt{N}\, \Sigma\, ,\quad
N\delta = t\, ,
\end{equation}
(\ref{seff2nd}) reduces to,
\begin{equation}
\label{Ssca2}
S_{\rm scaled} = -{1\over 8\pi} \int_{\Lambda_{0}}\! d^{2}\sigma \left[
{1\over 4 |\Lambda|^{2}}\int\! d^{4}\Theta \, \bar S S + \re\int\!
d\Theta_{-} d\bar\Theta_{+} \Bigl( t S - {S^{3}\over 
3\Lambda^{2}}\Bigr)\right] .
\end{equation}
No remormalization is needed for the action (\ref{Ssca2}), and thus 
the cut-off $\Lambda_{0} = \sqrt{N}|\Lambda|$ can be removed. 
The double scaled theory for a second class 
singularity is thus given by a simple Landau-Ginzburg action, as was 
suggested in Section 3.3.2. It coincides at low energy with the first 
class double scaled theory (\ref{lscaeff}), but differs from it in the UV.

\section{${\cal N}=1$ supersymmetric models}
\setcounter{equation}{0}

The purpose of the following brief Section is to emphasize the fact that 
the results obtained so far do not depend on 
supersymmetry. We focused on a supersymmetric model because our main 
goal was to make a comparison with the four dimensional supersymmetric gauge 
theories studied in \cite{fer1,fer2}. In fact, we believe that 
constructions of non-supersymmetric four dimensional non-critical 
strings could be made by using non-supersymmetric gauge theories with 
Higgs fields and adjusting the parameters in the Higgs potential to 
approach a critical point. 

The ${\cal N}=2$ supersymmetric model we have studied in
details in this paper was based on the
classical potential (\ref{Vcl}) with the constraints 
(\ref{cpNeq}, \ref{cpNid}) on the complex fields $\phi_{i}$. It is 
natural to suspect that a very similar, but only ${\cal N}=1$ supersymmetric, 
model could be constructed for which the classical potential is
\begin{equation}
V_{\rm cl}={1\over 2}\sum_{i=1}^{N+1}(\sigma + m_{i})^{2}\,\phi_{i}^{2}\, ,
\end{equation}
with real fields $\phi_{i}$, $\sigma$ and mass parameters $m_{i}$,
and constraint
\begin{equation}
\label{spheremod}
\sum_{i=1}^{N+1} \phi_{i}^{2} = {4\pi\over g^{2}}\,\cdotp
\end{equation}
Such a model indeed exists. By introducing Majorana 
spinors $\psi_{i}$ which are in the same supermultiplet as the 
$\phi_{i}$s and a supermultiplet $(\mu , D)$ of Lagrange multipliers, 
its lagrangian reads
\begin{eqnarray}
&&\hskip -1.5cm L = \sum_{i=1}^{N+1} \Bigl( {1\over 2}\, \phi_{i}\left(
\partial^{2} - (\sigma + m_{i})^{2}\right)\phi_{i}^{2}
+ \bar\psi_{i} \left( i\partialsl + \sigma + 
m_{i}\strut\right)\psi_{i}\Bigr)\nonumber\\
&&\hskip 5cm +{D\over 2}\, \Bigl( \sum_{i=1}^{N+1}\phi_{i}^{2} - {4\pi\over 
g^{2}}\Bigr) - \bar\mu\sum_{i=1}^{N+1}\psi_{i}\phi_{i}\, 
.\label{LN1}\\ \nonumber
\end{eqnarray}
This is a $S^{N}$ non-linear $\sigma$ model with ${\cal N}=1$ 
supersymmetry. The mass terms come from a superpotential
\begin{equation}
\label{WN1}
W(\Phi_{i}) = {1\over 2} \sum_{i=1}^{N+1} m_{i}\Phi_{i}^{2}\, .
\end{equation}
The lagrangian (\ref{LN1}) is very similar 
to (\ref{lag}), and the large $N$ limit can be studied with the same 
methods \cite{fer5}. A particularly interesting aspect of the model 
(\ref{LN1}) is that the number of vacua changes when the mass 
parameters are varied, while the Witten index $\tr (-1)^{F} = 1 + (-1)^{N}$
is, of course, constant. In the $\cpN$ model, holomorphicity was preventing
such drastic changes in the space of vacua. At large $N$, 
one can show that (\ref{vac}) is replaced by 
\begin{equation}
\label{vacN1}
\prod_{i=1}^{N+1} (\sigma + m_{i})^{2} = \Lambda^{2(N+1)}\, ,
\end{equation}
where all the parameters are now real. In the $\Lambda\rightarrow 0$
weak coupling limit, we have $2(N+1)$ vacua coming in $N+1$  
inequivalent pairs,
\begin{equation}
\label{wcN1}
\langle\sigma\rangle _{\rm w.c.}\simeq -m_{i}\, ,\quad \langle\phi_{j}
\rangle_{\rm w.c.} \simeq \pm {2\sqrt{\pi}\delta_{ij}\over g}\,\cvp
\end{equation}
while at strong coupling we are left with only two vacua
\begin{equation}
\label{scN1}
\langle\sigma\rangle_{\rm s.c.} \simeq \pm\Lambda\, .
\end{equation}
The particular values of the mass parameters for which the number of 
vacua changes correspond to critical points where the large $N$ 
expansion breaks down, and where double scaling limits can be defined. 
The double scaled theories are typically simple Landau-Ginzburg 
theories \cite{fer5}. The scalings always look like (\ref{sca1}), with 
logarithmic correction, because ${\cal N}=1$ Landau-Ginzburg models 
need to be renormalized by normal ordering the superpotential.

Instead of considering superpotential-induced mass terms,
one can also use the isometries of the target space, as was 
done for the $\cpN$ model. In that case, the ${\cal N}=1$ 
supersymmetric lagrangian takes the form 
\begin{eqnarray}
&&\hskip -1.5cm L = \sum_{k,l=1}^{N+1} \Bigl( {1\over 2}\, 
\phi_{k}\left(
(\partial^{2}-\sigma^{2})\delta_{kl} - \smash{\sum_{i=1}^{N+1}}
m_{ik}m_{il}\right)\phi_{l}^{2}
+ \bar\psi_{k} \left( (i\partialsl + \sigma)\delta_{kl} + 
m_{kl}\sigma^{3}\right)\psi_{l}\Bigr)\nonumber\\
&&\hskip 6cm +{D\over 2}\, \Bigl( \sum_{i=1}^{N+1}\phi_{i}^{2} - {4\pi\over 
g^{2}}\Bigr) - \bar\mu\sum_{i=1}^{N+1}\psi_{i}\phi_{i}\, ,
\label{LN2}\\ \nonumber
\end{eqnarray}
where $m_{ij}$ is an antisymmetric mass matrix.
Again, critical points can be found, and double scaling limits 
defined \cite{fer5}. An interesting aspect of the model (\ref{LN2}) is that 
supersymmetry can be spontaneously broken when $N$ is odd.

In both models (\ref{LN1}) and (\ref{LN2}), the masses 
induce a quadradic term $h_{ij}\phi_{i}\phi_{j}/2$, where the tensor
``magnetic field'' ${\mathbf h} = \mathbf m \mathbf m^{T}$ is expressed in 
term of a symmetric or antisymmetric mass matrix respectively.
One cannot find a 
supersymmetric theory with arbitrary $\mathbf h$, but the 
corresponding non-supersymmetric model can also be studied, and again
critical points are found and double scaling limits can be defined
\cite{fer3}.

\section{Prospects}
\setcounter{equation}{0}
The main goal of the present paper was to improve one's understanding of 
the results obtained in \cite{fer1,fer2}. We hope that our analysis 
has convinced the reader that the gauge theories double 
scaling limits are likely to yield well-defined four dimensional
non-critical
string theories, as conjectured in \cite{fer2}. A natural avenue for 
future work is to try to generalize the cases studied in \cite{fer2}.
There is a variety of critical points appearing on 
the moduli space of supersymmetric gauge theories, and
a large class of string theories can certainly be generated. 
Remarkably, for all those theories, the dimensions of the world sheet 
couplings as well as the space-time central charge as a function of these 
couplings can be calculated exactly. It would be interesting to study
in details the structure of the formulas so obtained.
One may hope, from experience with the $c<1$ matrix models where 
the KdV hierarchy plays a prominent r\^ole \cite{review2}, that a general 
mathematical structure could emerge. Unravelling this structure 
might eventually help in understanding our string theories from a 
more conventional, `continuous' point of view.

Another fascinating possible direction of research is based on the 
fact, explained in the Introduction, that our two dimensional models 
admit brane constructions. A crucial feature
is that the large $N$ limit of the $\sigma$ models correspond 
to a large number of branes, as in the case of gauge theories \cite{malda}.
One might then expect to be able to find a description involving 
quantum gravity. The 
startling point is that, in sharp contrast with the gauge theory case, the 
large $N$ limits of our models are exactly solvable.
\section*{Acknowledgements}
I have enjoyed 
useful discussions with Igor Klebanov, Nikita Nekrasov and
Sasha Polyakov. This work was supported in part by a Robert H.~Dicke 
fellowship and by the Swiss National Science Foundation.

\begin{appendix}
\section{Formulas for determinants}
\setcounter{equation}{0}
We consider the renormalized euclidean determinants
\begin{equation}
\label{defdet}
\det_{\rm r} (-\nabla^{2} + h) = \exp (2\,\xi[h,v_{\alpha}])\, ,\quad
\det_{\rm r} (\nablasl - \check f) =\exp (2\, \zeta[f,v_{\alpha}])\, ,
\end{equation}
where the functionals $\xi$ and $\zeta$ are defined by the equations
\begin{eqnarray}
\label{funcdef}
&&\xi[h,v_{\alpha}]= {1\over 2} \tr\ln (-\nabla^{2} + h) - 
\xi_{0}[h,v_{\alpha}] \, ,\label{sdef}\\
&&\zeta[f,v_{\alpha}] = {1\over 2}\tr\ln (\nablasl - \check f) - 
\zeta_{0}[f,v_{\alpha}]\, ,\label{zetdef}\\ \nonumber
\end{eqnarray}
with local counterterms $\xi_{0}$ and $\zeta_{0}$. The covariant 
derivative is $\nabla_{\alpha} = \partial_{\alpha} + i
v_{\alpha}$, $h$ is a real and positive field, and $f$
is a complex field with associated matrix $\check f = \re f - 
i \sigma^{3}\im f $. The local counterterms depend on the 
regularization and renormalization schemes. Gauge invariant and 
supersymmetric results can be obtained by using a Pauli-Villars 
regularization with cut-off $\Lambda_{0}$, renormalization scale 
$\mu$ and counterterms
\begin{equation}
\label{ctdef}
\xi_{0} = {1\over 4\pi}\ln {\Lambda_{0}\over\mu}\, \int\! d^{2}x\, h\, ,\quad
\zeta_{0} = {1\over 4\pi} \ln{\Lambda_{0}\over\mu}\, \int\! d^{2}x\, 
|f|^{2}\, .
\end{equation}
Dimensional regularization $D=2-\epsilon$ can also be used, with the 
definition
\begin{equation}
\label{dregdef}
{1\over 2\pi}\ln {\Lambda_{0}\over\mu} = \int\! {d^{D}p\over (2\pi)^{D}}\, 
{1\over p^{2}+\mu^{2}} = {1\over 2\pi\epsilon} - {1\over 4\pi} \ln 
{\mu^{2}\over 4\pi} - {\gamma\over 4\pi}\, \cvp
\end{equation}
but one 
must add a finite local counterterm $-\int d^{2}x (\im f)^{2}/(4\pi )$ 
to the fermionic functional for supersymmetry to be preserved (such a 
term is generated due to the unusual properties of the 
$\bar\psi\sigma^{3}\psi \im f$ vertex in dimensional regularization).
\subsection{Special cases}
For $v_{\alpha}$ pure gauge, $h$ constant and positive, and $f$ constant, 
we have
\begin{equation}
\label{potfor}
\hskip -.2cm\xi[h={\rm cst}>0,v_{\alpha}=\partial_{\alpha}\chi] 
= \int\! d^{2}x\, V_{\rm b} (h)\, ,\quad
\zeta[f={\rm cst},v_{\alpha}=\partial_{\alpha}\chi] = \int\! d^{2}x\,
V_{\rm f}(f)\, ,
\end{equation}
with the potentials
\begin{equation}
\label{pot}
V_{\rm b}(h) = -{h\over 8\pi } \ln {h\over e\mu^{2}}\, \cvp\quad
V_{\rm f}(f) = V_{\rm b}(|f|^{2})\, .
\end{equation}
If $f=0$, we have, for any $v_{\alpha\beta}$ going to zero fast 
enough at infinity ,
\begin{equation}
\label{zetasf}
\zeta[f=0,v_{\alpha}] = {1\over 8\pi}\int\! d^{2}x\, 
v_{\alpha\beta}\,{1\over \partial^{2}}\,v_{\alpha\beta}\, .
\end{equation}
If $\im f/\re f$ is constant, the term linear in $v_{\alpha\beta}$
in $\zeta$ can also be exactly calculated,
\begin{equation}
\label{linv}
\zeta[f,v_{\alpha}]_{{\rm linear\ in\ }\tilde v} = {i\over 8\pi}\int\! 
d^{2}x\, \epsilon_{\alpha\beta}v_{\alpha\beta}
\, \im\ln f\, \cdotp
\end{equation}
\subsection{General case}
In general, one writes
\begin{equation}
\label{expf}
h = m^{2} + \varphi\, ,\quad f = M + \phi\, ,
\end{equation}
with $m$ real and $M$ complex, and one expands the 
functionals in powers of the fields $\varphi$, $\phi$ and 
$v_{\alpha}$,
\begin{equation}
\label{exppower}
\xi[h,v_{\alpha}] = \sum_{n=0}^{\infty} \xi_{n}[\varphi 
,v_{\alpha};m^{2}]\, ,\quad \zeta[f,v_{\alpha}] = 
\sum_{n=0}^{\infty} \zeta_{n}[\phi,v_{\alpha};M]\, .
\end{equation}
The functionals 
\begin{eqnarray}
\label{linfunc}
&&\xi_{1}= {1\over 8\pi}\ln {\mu^{2}\over m^{2}} \int\! d^{2}x\, 
\varphi\, ,\label{s1}\\
&&\zeta_{1} = {i\over 8\pi} \int\! d^{2}x\, 
\epsilon_{\alpha\beta}v_{\alpha\beta} \im\ln M + 
{1\over 2\pi} \ln {\mu\over |M|}\int\! d^{2}x\, \re (M^{*}\phi)\, 
,\label{z1}\\ \nonumber
\end{eqnarray}
are linear in the fields, $s_{2}$ and $\zeta_{2}$ are quadratic, etc\ldots 
The quadradic pieces are most easily expressed by introducing the 
Fourier transforms of the fields,
\begin{equation}
\label{Fourierdef}
\hat v_{\alpha} (k) = \int\! d^{2}x\, e^{-ikx}v_{\alpha}(x)\, ,\quad {\rm 
etc,}
\end{equation}
and the one-loop integral
\begin{equation}
\label{Sdef}
S(k;m^{2}) = -{1\over 8\pi^{2}} \int {d^{2}p\over 
(p^{2}+m^{2})((p+k)^{2}+m^{2})}\, \cdotp
\end{equation}
This integral can be evaluated in different regimes (euclidean, and 
below or above the pair production threshold), by using Feynman's 
$i\epsilon$ prescription when necessary,
\begin{equation}
\label{Svalue}
S(k;m^{2}) = \left\{\matrix{
-{\strut\displaystyle 1\over\strut\displaystyle 4\pi}\, {\strut
\displaystyle 1\over 
\strut\displaystyle p^{2}\sqrt{1+4m^{2}/p^{2}}}\, 
\ln {\strut\displaystyle \sqrt{1+4m^{2}/p^{2}} + 1\over\strut\displaystyle 
\sqrt{1+4m^{2}/p^{2}} 
-1}\, \cvp\ {\rm for\ } p^2 >0\, ,\hfill\cr\noalign{\smallskip}
{\strut\displaystyle 1\over\strut\displaystyle 2\pi}\,
{\strut\displaystyle 1\over\strut\displaystyle p^{2}\sqrt{-1-4m^{2}/p^{2}}}
\, \arctan {\strut\displaystyle 1\over\strut
\displaystyle\sqrt{-1-4m^{2}/p^{2}}}\,\cvp\ {\rm 
for\ } 0< -p^{2}< 4m^{2}\, ,\hfill\cr\noalign{\smallskip}
-{\strut\displaystyle 1\over\strut\displaystyle 4\pi}\, {\strut
\displaystyle 1\over\strut
\displaystyle p^{2}\sqrt{1+4m^{2}/p^{2}}}\,\left( 
\ln {\strut\displaystyle 1+ \sqrt{1+4m^{2}/p^{2}}\over\strut\displaystyle
1- \sqrt{1+4m^{2}/p^{2}}} - i\pi \right) 
\!,\ {\rm for\ } -p^2 > 4 m^2 .\hfill\cr} \right.  
\end{equation}
Introducing $\phi_{1}(x) = \re\phi(x)$, $\phi_{2}(x)=\im\phi(x)$, 
$M_{1}=\re M$ and $M_{2}=\im M$, we have 
\begin{eqnarray}
\label{quadfunc}
&&\hskip -1.2cm  \xi_{2} = {1\over 2}\int\! {d^{2}k\over (2\pi)^{2}} \Biggl[
\Bigl( (k^{2}+4m^{2})S(k;m^{2}) + {1\over 2\pi}\Bigr)\Bigl(
{k_{\alpha}k_{\beta}\over k^{2}} - \delta_{\alpha\beta}\Bigr) \hat 
v_{\alpha}(-k)\hat v_{\beta}(k)  \nonumber \\ 
&&\hskip 7.75cm + S(k;m^{2}) \hat\varphi (-k)\hat\varphi (k)\Biggr]\,
,\label{s2}\\
&&\hskip -1.2cm \zeta_{2} = {1\over 2}\int\! {d^{2}k\over (2\pi)^{2}}
\Biggl[ \Bigl( 4|M|^{2} S(k;|M|^{2}) + {1\over 2\pi} \Bigr)
\Bigl( {k_{\alpha}k_{\beta}\over k^{2}} - \delta_{\alpha\beta}\Bigr) \hat 
v_{\alpha}(-k)\hat v_{\beta}(k)  \nonumber\\
&&\hskip 2.5cm + \Bigl( (k^{2}+4M_{1}^{2})S(k;|M|^{2}) + {1\over 2\pi}\ln 
{\mu\over |M|}\Bigr) \hat\phi_{1}(-k)\hat\phi_{1}(k) \nonumber\\
&&\hskip 2.5cm + \Bigl( (k^{2}+4M_{2}^{2})S(k;|M|^{2}) + {1\over 2\pi}\ln 
{\mu\over |M|}\Bigr) \hat\phi_{2}(-k)\hat\phi_{2}(k)  \nonumber\\
&&\hskip 2.5cm +4 S(k;|M|^{2})\epsilon_{\alpha\beta} k_{\beta}
\, \hat v_{\alpha}(-k) (M_{1}\hat\phi_{2}(k) - 
M_{2}\hat\phi_{1}(k)) \nonumber \\
&&\hskip 2.5cm + 8M_{1}M_{2} S(k;|M|^{2})\, \hat\phi_{1}(-k)
\hat\phi_{2}(k)\Biggr].\label{z2} \\ \nonumber
\end{eqnarray}
The low energy expansion up to two derivative terms is obtained by using
\begin{equation}
\label{squadlp}
S(k;m^{2}) = -{1\over 8\pi m^{2}}\,\Bigl( 1-{k^{2}\over
6m^{2}}+{\cal O}(k^{4})\Bigr)
\end{equation}
and it reads
\vskip -.5cm\
\begin{eqnarray}
&&\hskip -1.5cm \xi_{2}^{\rm l. e.} =
{1\over 48\pi m^{2}}\int\! d^{2}x\, \Biggl[ {1\over 2} 
v_{\alpha\beta}v_{\alpha\beta} + {1\over 
2 m^{2}}\partial_{\alpha}\varphi\partial_{\alpha}\varphi - 
3\varphi^{2}\Biggr] ,\label{les2} \\
&&\hskip -1.5cm \zeta_{2}^{\rm l. e.} =
{1\over 48\pi |M|^{2}}\int\! d^{2}x\, \Biggl[
- v_{\alpha\beta}v_{\alpha\beta} + 6i \im (M^{*}\phi) 
\epsilon_{\alpha\beta}v_{\alpha\beta} - 3 \,
\partial_{\alpha}\phi^{\dagger}\partial_{\alpha}\phi\nonumber\\
&&\hskip -.5cm +{2\over |M|^{2}} \re (M\partial_{\alpha}\phi) 
\re (M\partial_{\alpha}\phi)+ 12 |M\phi|^{2}\ln {\mu\over |M|} - 12
\left( \re (M^{*}\phi)\strut \right)^{2}\Biggr].\label{lez2}\\ \nonumber
\end{eqnarray}
\end{appendix}
\vskip -2cm\
\end{document}